\documentclass{aa}
\usepackage{graphicx,natbib}
\usepackage[varg]{txfonts}
\usepackage{float}
\usepackage{afterpage}
\usepackage{color}
\usepackage{upgreek}
\usepackage{soul}
\usepackage{hyperref}
\usepackage{lineno}
\newcommand{\red}{\textcolor{red}}

\newcommand{\Porb}{\ensuremath{P_{\mathrm{orb}}}}
\newcommand{\Prot}{\ensuremath{P_{\mathrm{rot}}}}
\newcommand{\Bm}{\ensuremath{\langle B\rangle}}
\newcommand{\Bz}{\ensuremath{\langle B_z\rangle}}

\newcommand{\vsi}{\ensuremath{v\,\sin i}}

\newcommand{\kms}{km\,s$^{-1}$}
\newcommand{\Feline}{Fe~\textsc{ii}~$\uplambda\,6149.2$\,\AA}

\newcommand{\vr}{\ensuremath{v_{\mathrm{r}}}}
\newcommand{\fo}{\phantom{1}}

\defcitealias{1997A&AS..123..353M}{M97}
\defcitealias{2017A&A...601A..14M}{M17}
\defcitealias{2022MNRAS.514.3485G}{G22}

\bibpunct[ ]{(}{)}{;}{a}{}{,}

\begin{document}

\title{Ap stars with resolved magnetically split lines:\\
 Mean magnetic field modulus measurements 1999--2004} 

\author{G.~Mathys\inst{1}
  \and J.~D.~Landstreet\inst{2,3}
  \and T.~A.~A.~Sigut\inst{3,4}}  

\institute{European Southern Observatory,
  Alonso de Cordova 3107, Vitacura, Santiago, Chile\\\email{gmathys@eso.org}
  \and
Armagh Observatory and Planetarium, College Hill, Armagh, BT61 9DG,
Northern Ireland, UK
\and
Department of Physics \& Astronomy, University of
Western Ontario, 1151 Richmond Street, London, Ontario, N6A 3K7,
Canada 
\and
Institute for Earth and Space Exploration (IESX), The University of
Western Ontario, 7134 Perth Drive, London, ON N6A 5B7, Canada} 

\date{Received $\ldots$ / Accepted $\ldots$}

\titlerunning{Mean magnetic field modulus measurements 1999--2004}

\abstract
{Magnetic Ap stars that have a strong enough field and a sufficiently
  low $\vsi$ show spectral lines resolved into their magnetically
  split components. This allows their mean magnetic field modulus
  $\Bm$ to be determined. }
{We complement the existing knowledge of the rotational magnetic
  variability and of the binarity of a sample of 34 Ap stars with
  resolved magnetically split lines by analysing observations obtained
  between 1999 and 2004 that had not been exploited before and, when
  available, observatory archive data.}
{We determine the values of $\Bm$ and of the radial velocity by
  measuring the magnetic splitting and the wavelength shift of the
  \Feline\ line in high-resolution spectra recorded in natural
  light. Combining our measurements with published ones, we
  characterise the variability of both quantities.}
{We derived for the first time the accurate, unambiguous values of the
  rotation periods of HD~29578 and HD~165474. We set lower limits of
  years to decades to
  the rotation periods of six stars and identified four more that
  have not shown any significant $\Bm$ variations over timescales of
  13 to 33 years. We established that HD~47103 is the first Ap star
  with a rotation period of years known to have a magnetic field
  significantly exceeding 7.5\,kG. We achieved the first determination
  of the orbital elements of HD~116114, HD~142070, and HD~144897, and 
  we refined the orbital elements of six other binaries.}
{}


\maketitle
\nolinenumbers

\section{Introduction}
\label{sec:intro}
Eight decades have elapsed since \citet{1947ApJ...105..105B}
discovered the first magnetic field in a star other than the Sun. This
star, 78~Vir (= HD~118022), belongs to the group of the main sequence
early F to mid-B stars that show large surface abundance anomalies and
inhomogeneities, commonly referred to as Ap stars. Their large-scale
organised magnetic fields, whose strengths range from a few 10\,G to a
few 10\,kG, have structures that, to first order,
resemble a dipole inclined to the stellar rotation axis. In most cases,
these fields are diagnosed through spectropolarimetry, since
the Zeeman splitting of the spectral lines is smeared out by
rotational Doppler broadening. However, several percent of the Ap
stars have a low enough \vsi\ and a magnetic field strong enough so
that resolved magnetically split lines (RMSL) can be observed in their
spectrum. Such cases present particular interest as they allow the
magnetic field to be characterised in a mostly model-free and
approximation-free manner, with only limited dependence on the
geometry of the observation. 

\citet{1960ApJ...132..521B} reported the first observation of RMSL in
a star other than the Sun: the B9p star HD~215441
(=~BD+54~2846). \citet {1971ApJ...164..309P} carried out the first 
systematic study of stars having this property, of which he listed
seven. For the next 20 years, while Ap stars showing magnetically
resolved lines were occasionally discovered, they remained little
studied isolated specimens. It is only around 1990 that the possible
existence of of a substantial population of stars of this type was
realised \citep{1990A&A...232..151M,1992A&A...256..169M}. This
triggered extensive searches for Ap stars with RMSL
\citep[; the first of these papers is hereafter referred to as
M97]{1997A&AS..123..353M,2008MNRAS.389..441F,2012MNRAS.420.2727E,2019ApJ...873L...5C}  
and systematic follow-up studies of their properties
\citep{2017A&A...601A..14M,2022MNRAS.514.3485G}. These two papers are
referred to below as M17 and G22, respectively.

The datasets analysed in the latter two references are extensive. The
high resolution Stokes $I$ spectra of 43 stars analysed by
\citetalias{2017A&A...601A..14M} span the time interval from 1988 May to
1998 September. The bulk of the data for the 36 stars studied by
\citetalias{2022MNRAS.514.3485G} were obtained between 2001 and 2021.

We recently recovered a collection of 103 high-resolution spectra of
Ap stars 
acquired between 1999 April and 2004 July with the Coud\'e Echelle
Spectrograph (CES) at the European
Southern Observatory (ESO) and with the Gecko coud\'e spectrograph of
the Canada-France-Hawaii Telescope (CFHT). Of these, 82
show RMSL; they pertain to 39 different stars.
Most of these spectra had been obtained at
a time before the start of systematic archival of observations in real
time by observatories, and we had lost track of them. The measurements
obtained from their analyses bridge gaps between the time series of
data presented in M17 and G22, and accordingly
provide additional constraints on the variations of the studied
stars. This makes their publication valuable.

The 1999--2004 CES or Gecko spectra of four of the above-mentioned 39
stars with RMSL have previously been
included in detailed individual studies: \citet[; HD~18078, 2
spectra]{2016A&A...586A..85M}, \citet[; HD~50169, 2
spectra]{2019A&A...624A..32M}, \citet[; HD~965, 4
spectra]{2019A&A...629A..39M}, and \citet[; HD~166473, 7
spectra]{2020A&A...636A...6M}. We do not repeat here the analysis of
these 15 spectra. Furthermore, three of the Gecko spectra showing
RMSL pertain to HD~201601 (= $\gamma$~Equ),
one of the best studied Ap stars. These spectra have already been
analysed by \citetalias{2022MNRAS.514.3485G}. The values of
the mean magnetic field modulus $\Bm$ (the line intensity weighted
average over the visible stellar disk of the modulus of the magnetic
vector) that they derived do not significantly differ from ours. They 
represent only a minor addition to the extensive set of $\Bm$ values
that have been accumulated for this star over many years, which is
exhaustively presented by \citetalias{2022MNRAS.514.3485G}. Repeating
here the discussion of these measurements would be
pointless. Accordingly, our new spectra of HD~201601 are also omitted
from 
the present study. Finally, HD~9996 (HR~465) shows magnetically resolved
lines only over a fraction of its 21.7\,yr rotation period
\citepalias{2017A&A...601A..14M}. The two new Gecko spectra acquired in
2000 July and 2004 July do not show any hint of resolved magnetic line
splitting, so that we do not include them in the present study, even
though they were considered by \citetalias{2022MNRAS.514.3485G}.

These observations from 1999--2004 and their analysis are described in
Appendix~\ref{sec:obs}. The mean magnetic field modulus
determinations are presented in 
Sect.~\ref{sec:meas} on a star-by-star basis. The spectral type
indicated in the subsection header for each star is from the `Renson
catalogue' \citep{2009A&A...498..961R}. The radial velocity (RV)
measurements derived as a by-product of our analysis and their
contribution to the study of the spectroscopic binaries of our
sample are discussed in Appendix~\ref{sec:SB}. An overview of the
obtained results and their implications for our knowledge of Ap stars
is given in Sect.~\ref{sec:disc}.

\section{Mean magnetic field modulus}
\label{sec:meas}

\subsection{HD~2453 (BD+31~59; A1p SrEuCr)}
\label{sec:hd_2453}
Our only new Gecko spectrum of HD~2453 was also analysed by
\citetalias{2022MNRAS.514.3485G}. The value that we derive for $\Bm$
does not significantly differ from the one that they report. 

\begin{figure}
  \centering
  \includegraphics[width=1.0\linewidth,angle=0]{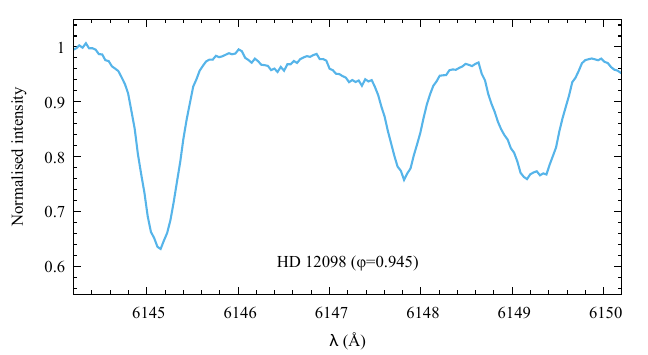}
  \caption{Portion of a CFHT Archive ESPaDOnS spectrum of
    HD~12098. The three main lines are Nd~{\sc
      iii}~$\lambda$\,6145.1\,\AA, Fe~{\sc ii}~$\lambda$\,6147.7\,\AA,
    and \Feline. Their Zeeman patterns are shown in Fig.~3 of
    \cite{2024A&A...691A.186M}. The \Feline\ doublet is marginally
    resolved into its magnetically split components.}
\label{fig:hd12098_6149}
\end{figure}

\begin{figure}
  \centering
  \includegraphics[width=1.0\linewidth,angle=0]{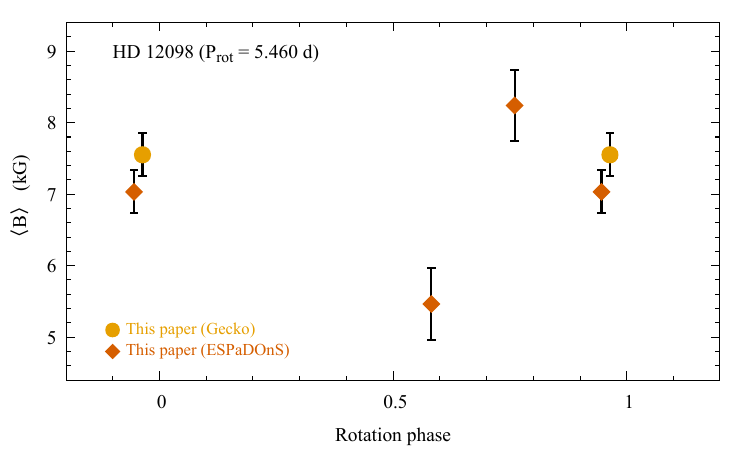}
\caption{Mean magnetic field modulus of HD~12098
  against rotation phase 
  computed with the value $\Prot=5\fd460$ of the rotation period
  and JD~2\,451\,889.42 as the phase origin
  \citep{2005A&A...429L..55R}. The latter is close to a positive
  extremum of the mean longitudinal magnetic field $\Bz$. Different 
  point types and colours are used to distinguish the observations
  from different origins.}
\label{fig:bm_phi_12098}
\end{figure}

\subsection{HD~12098 (BD+57~447; F0p Eu)}
\label{sec:hd_12098}
The presence of RMSL in the spectrum of HD~12098 is
reported here for the first time. That this resolution, 
for the \Feline\ line, is only marginal (see
Fig.~\ref{fig:hd12098_6149}) reflects the fact that the 
star has a short rotation period, $\Prot=5.460\pm0.001$\,d
\citep{2005A&A...429L..55R,2024MNRAS.529..556K}, so that the
Doppler broadening of the split line components is not negligible with
respect to their separation. The resulting line profile distortion
and the incomplete splitting resolution make it difficult to achieve
good precision in the determination of the mean magnetic field
modulus. The derived values are plotted against rotation phase in
Fig.~\ref{fig:bm_phi_12098}. The variation curve of the mean
longitudinal magnetic field $\Bz$ (that is, the line intensity
weighted average over the visible stellar disk of the component of the
magnetic vector along the line of sight) was established by
\citet{2005A&A...429L..55R} from observations at 17 epochs. 
With the time elapsed between these epochs and those of
our $\Bm$ determinations, the 
uncertainty of the relative phasing between the variations of the two
field moments is only of the order of 0.05 cycle. However, the large
uncertainties affecting the $\Bm$ values do not allow firm conclusions
to be derived about the phase relation between the $\Bz$ and $\Bm$
variations. One can only note the good agreement between the two $\Bm$
values obtained close to phase 0 (respectively from a Gecko spectrum
and from an ESPaDOnS spectrum), which is consistent with the published
period. 

\subsection{HD~12288 (BD+68~144; A2p CrSi)}
\label{sec:hd_12288}
As in the case of HD~2453, the $\Bm$ measurement of our single
additional Gecko spectrum of HD~12288 yields a value that does not
significantly differ from the one reported by
\citetalias{2022MNRAS.514.3485G}.

\begin{figure}
  \centering
  \includegraphics[width=1.0\linewidth,angle=0]{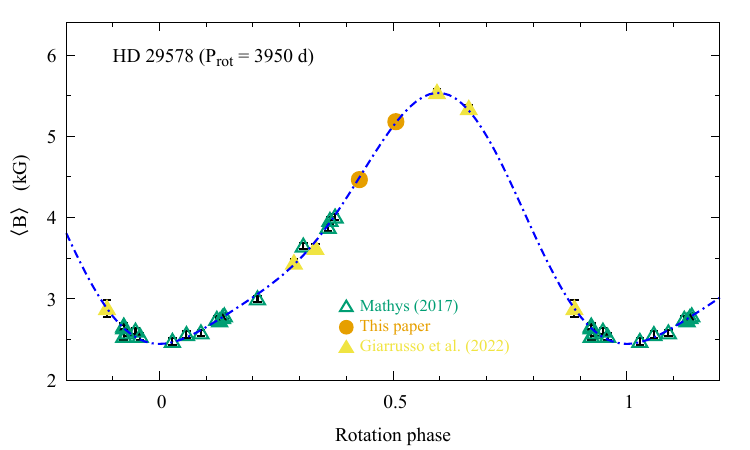}
\caption{Mean magnetic field modulus of HD~29578
  against rotation phase 
  computed with the value $\Prot=3950$\,d of the rotation period
  derived here. The phase origin has been
  set at JD~2\,449\,600.0, which is close to a $\Bm$ minimum. The
  dash-dotted blue curve represents the 
  least-squares fit of the measurements by a cosine wave and its first
  harmonic. Different 
  point types and colours are used to distinguish the observations
  from different origins.}
\label{fig:bm_phi_29578}
\end{figure}

\subsection{HD~29578 (CPD$-$54~685; A4p SrEuCr)}
\label{sec:hd_29578}
Besides our mean magnetic field modulus measurements and those of
\citetalias{2022MNRAS.514.3485G}, we include in our analysis a $\Bm$ value
derived by \citet{2004A&A...423..705R} from a CES spectrum recorded in
2001. The period search that we carried out using this dataset yields an
unambiguous result: $\Prot=3950\pm75$\,d. The variation of the mean
magnetic field modulus with this period is shown in
Fig.~\ref{fig:bm_phi_29578}. The time span covered by the analysed
observations, 7758 days, is only marginally shorter than two full
cycles. This allowed us to evaluate the uncertainty of the derived
period value by
plotting similar phase diagrams with series of progressively longer
and shorter periods, to constrain the range within which measurements
obtained at similar phases during each cycle remain consistent with
each other.

The 3950\,d value that we derive for the rotation period does not
significantly differ from one of the values proposed by
\citetalias{2022MNRAS.514.3485G} (4000\,d). By contrast, the alternative value
mentioned by these authors, 9370\,d, is definitely ruled out by our
analysis. Actually, such a value represents an extrapolation, as it
exceeds by a factor 1.2 the time interval spanned by the available data.

The $\Bm$ variation curve is strongly anharmonic and the
ratio between its extrema, $q\sim2.2$, is exceptionally large
\citepalias[see the distribution of the $q$ values in Fig.~3
of][]{2017A&A...601A..14M}. This 
indicates that the structure of the magnetic field of HD~29578
strongly departs from a dipole. In particular, the lack of symmetry of
the $\Bm$ variation curve shows that the field is not axisymmetric.

\begin{figure}
  \centering
  \includegraphics[width=1.0\linewidth,angle=0]{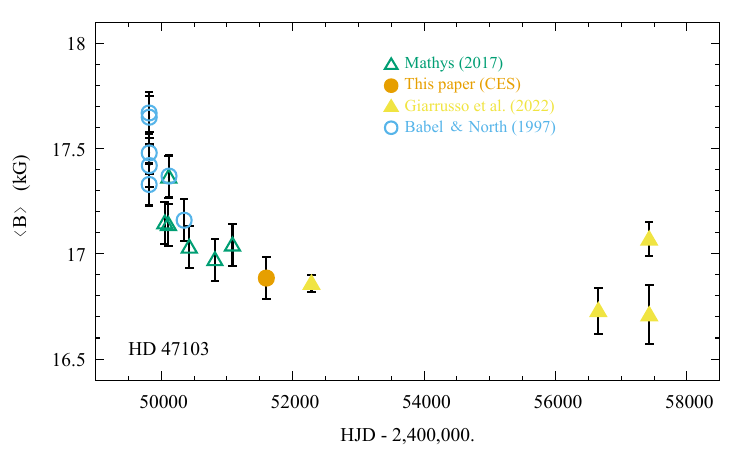}
\caption{Mean magnetic field modulus of HD~47103
  against Heliocentric Julian Date. Different
  point types and colours are used to distinguish the observations
  from different origins.}
\label{fig:bm_hjd_47103}
\end{figure}

\subsection{HD~47103 (BD+20~1508; Ap SrEu)}
\label{sec:hd_47103}
\citet{1995A&A...303L...5B} discovered the very strong magnetic field
($\Bm\sim17.5$\,kG) of HD~47103. Determinations of the mean magnetic
field modulus were also reported by \citet{1997A&A...325..195B}. We
combined the $\Bm$ values from these references with our own data and
the four new measurements of \citetalias{2022MNRAS.514.3485G}, and plotted
them against the date of observation in
Fig.~\ref{fig:bm_hjd_47103}. All the data from 1995 (the discovery
spectrum) to 2002 (the first measurement of
\citetalias{2022MNRAS.514.3485G}) draw a smooth monotonic variation
curve, suggestive of a rotation period much longer than the time range
covered by the observations: $\Prot\gg6.8$\,yr. Our new determination
of the mean magnetic field 
modulus unambiguously rules out the $17\fd683$ period proposed by
\citetalias{2022MNRAS.514.3485G}. According to 
the ephemeris of \citetalias{2022MNRAS.514.3485G}, the observation phase
would be $\phi=0.95$, rather close to the $\Bm$ maximum. By contrast,
the field value that we derive, $\Bm=16\,883$\,G, is among the
lowest ones for this star.

The three most recent measurements of \citetalias{2022MNRAS.514.3485G} show
somewhat more scatter, but this is mostly related to a single point
and may plausibly result from an unidentified observational
glitch. In particular, it seems quite possible that the mean magnetic
field modulus may have remained nearly constant from 2002
(JD~2\,452\,286) to 2016 (JD~2\,457\,426): HD~47103 may be another
example of the rather frequent trend for 
$\Bm$ variation curves to be nearly flat-bottomed over a rather
wide range of phases (see, for instance, the case of HD~187474 in
\citetalias{2017A&A...601A..14M}). If so, the rotation period of HD~47103
would be considerably 
longer than the time span of all the observations of this study:
$\Prot\gtrsim21$\,yr. However, given the large time interval
separating the last three measurements from the previous ones and the
resulting uncertainty on the behaviour of the star in this gap, we
prefer to adopt conservatively the value suggested above,
$\Prot\gg6.8$\,yr, which is based on the consideration of 15 $\Bm$
values with good time sampling. 

In any event, there is no doubt left that with a period 
of years and an average mean magnetic field modulus of
$\sim$17.2\,kG, HD~47103 represents the first confirmed exception 
to the conclusion proposed by \citetalias{2017A&A...601A..14M}, that
phase-averaged mean magnetic field modulus values greater than
$\sim$7.5\,kG are only found in Ap stars with rotation periods shorter
than $\sim$150\,d. However, as HD~47103 is the only exception
identified until now, it remains likely that the rate of
occurrence of extremely strong magnetic fields tends to be
significantly higher in stars with rotation periods of up to a few
months than in stars that rotate more slowly.

\begin{figure}
  \centering
  \includegraphics[width=1.0\linewidth,angle=0]{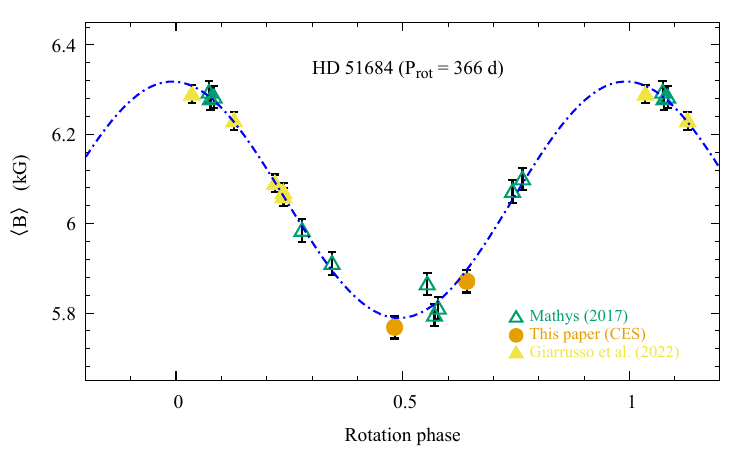}
\caption{Mean magnetic field modulus of HD~51684
  against rotation phase 
  computed with the value $\Prot=366$\,d of the rotation period
  \citepalias{2022MNRAS.514.3485G}. The phase origin has been
  set at JD~2\,449\,960.0, which is close to a $\Bm$ maximum. The
  dash-dotted blue curve represents the 
  least-squares fit of the measurements by a cosine wave. Different 
  point types and colours are used to distinguish the observations
  from different origins.}
\label{fig:bm_phi_51684}
\end{figure}

\subsection{HD~51684 (CD$-$40~2796; F0p SrEuCr)}
\label{sec:hd_51684}
As illustrated in Fig.~\ref{fig:bm_phi_51684}, the two $\Bm$ values
that we derived from additional CES spectra are fully consistent with
the refined value of the 
rotation period ($\Prot=366$\,d) determined by
\citetalias{2022MNRAS.514.3485G}. The two new measurements of the mean
magnetic field modulus reported here do not extend the time base
covered by the available data, but they valuably constrain the $\Bm$
minimum.

\begin{figure}
  \centering
  \includegraphics[width=1.0\linewidth,angle=0]{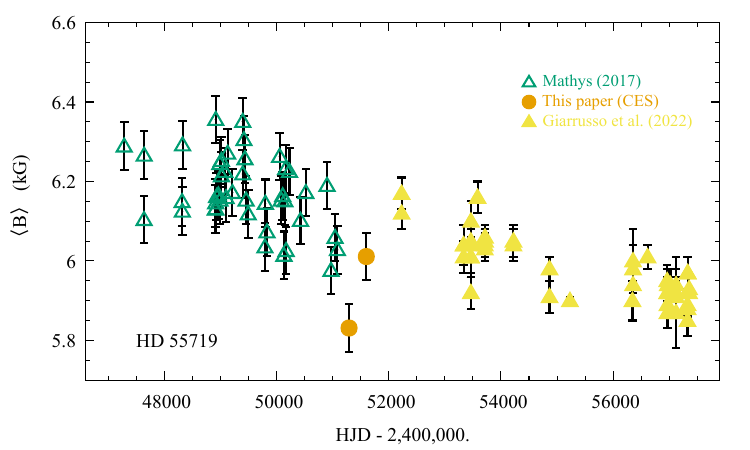}
\caption{Mean magnetic field modulus of HD~55719
  against Heliocentric Julian Date. Different
  point types and colours are used to distinguish the observations
  from different origins.}
\label{fig:bm_hjd_55719}
\end{figure}

\subsection{HD~55719 (HR~2727; A3p SrCrEu)}
\label{sec:hd_55719}
We add only two new $\Bm$ measurements (from CES spectra) of HD~55719 to the
data set studied by \citetalias{2022MNRAS.514.3485G}, but the revised values of
this magnetic field moment for the spectra considered by
\citetalias{2017A&A...601A..14M} tend to be systematically smaller than the
original ones. Accordingly, while the long-term trend of (mostly)
monotonic decrease of the field strength over nearly 30 years visible
in Fig.~13 of \citetalias{2022MNRAS.514.3485G} is confirmed, the
variation amplitude appears somewhat lower in
Fig.~\ref{fig:bm_hjd_55719}. Accordingly, we conservatively adopt
  a lower limit $\Prot\gg30$\,yr of the rotation
  period. (The higher lower limit given by
  \citetalias{2022MNRAS.514.3485G}, $\Prot>38$\,yr, is based on an
  extrapolated fit to an assumed sinusoidal $\Bm$ variation curve,
  which is poorly constrained.)

\subsection{HD~59435 (BD$-$8~1937; A4p SrCrEu)}
\label{sec:hd_59435}
The double-lined spectroscopic binary HD~59435 has been studied in
detail by \cite{1996A&A...314..491W,1999A&A...347..164W}. Its primary
component is a G8 or K0 giant; the secondary is an Ap star showing
RMSL. The orbital period of the system is
$\Porb=1386\fd1$ and the rotation period of the Ap star is
$\Prot=1360$\,d; the similarity of the two periods appears
coincidental. In general, the determination of the mean magnetic field
modulus of the Ap star requires disentangling the contributions of
the lines of the two components. Fortunately, the new CES spectrum
analysed here was acquired at orbital phase $\phi_{\rm orb}=0.843$, at
which the \Feline\ line of the Ap secondary is not significantly
affected by any line of the primary. This allowed us to determine
$\Bm$ in the same straightforward manner as for a single
star, yielding a value $\Bm=2683$\,G at rotation phase $\phi_{\rm
  rot}=0.191$. This value is reasonably
consistent with the variation curve of \citet{1999A&A...347..164W}.

\begin{figure}
  \centering
  \includegraphics[width=1.0\linewidth,angle=0]{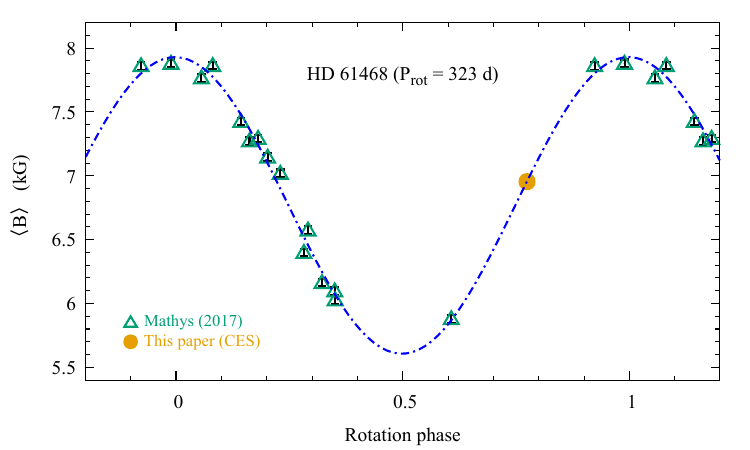}
\caption{Mean magnetic field modulus of HD~61468
  against rotation phase 
  computed with the value $\Prot=323$\,d of the rotation period
  derived here. The phase origin has been
  set at JD~2\,450\,058.5, which is close to a $\Bm$ maximum. The
  dash-dotted blue curve represents the 
  least-squares fit of the measurements by a cosine wave. Different 
  point types and colours are used to distinguish the observations
  from different origins.}
\label{fig:bm_phi_61468}
\end{figure}

\subsection{HD~61468 (CD$-$27~4341; A3p EuCr)}
\label{sec:hd_61468}
The additional $\Bm$ value derived from the CES spectrum acquired in
2000 represents a significant new constraint for determination
of the rotation period, as it fills a gap of more than 0.3 cycle in
the phase coverage (see Fig.~\ref{fig:bm_phi_61468}). The
revised value that we obtain, $\Prot=323\pm1$\,d, has a considerably
lower uncertainty than the value reported by
\citetalias{2017A&A...601A..14M}. The slightly different value adopted
by \citetalias{2022MNRAS.514.3485G} ($\Prot=321\pm1$\,d) is
constrained by a single $\Bm$ determination from an
additional HARPS-N spectrum. This $\Bm$ value is a definite outlier
with $\Prot=323$\,d. We do not know for sure why we cannot
reconcile it with our new CES measurement, but we note that
its uncertainty is well above the average for the study of
\citetalias{2022MNRAS.514.3485G}. Given that the \Feline\ line is not
affected by major blends and that it is very clearly split in
HD~61468, we suspect that this large uncertainty may reflect an
unusually low S/N  
of the analysed spectrum. This is why we give preference to the value
of the rotation period that we derive, even though it is based on
observations spanning a much shorter time base.

\subsection{HD~70331 (CD$-$47~3803; B8p Si)}
\label{sec:hd_70331}
The addition of a single CES spectrum to the data set analysed in
\citetalias{2017A&A...601A..14M} is insufficient to derive
significantly better constraints on the magnetic variability of
HD~70331. The most recent value of $\Bm$ remains within the range
spanned by those previously determined.

\begin{figure}
  \centering
  \includegraphics[width=1.0\linewidth,angle=0]{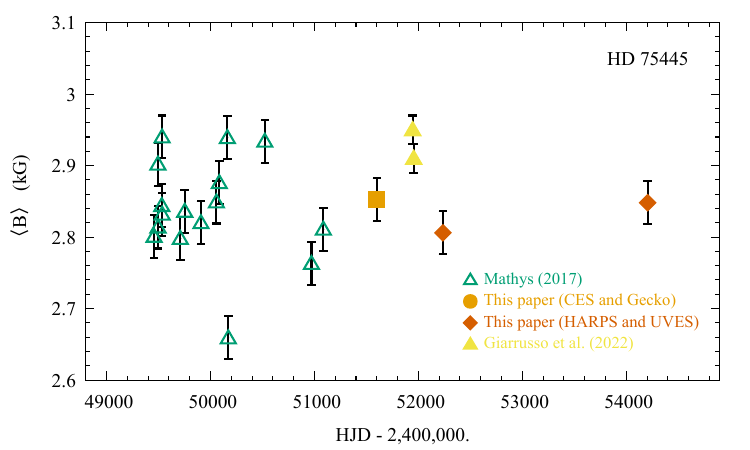}
\caption{Mean magnetic field modulus of HD~75445
  against Heliocentric Julian Date. Different
  point types and colours are used to distinguish the observations
  from different origins.}
\label{fig:bm_hjd_75445}
\end{figure}

\subsection{HD~75445 (CD$-$38~4907;A3p SrEu)}
\label{sec:hd_75445}
The slow, monotonic decline of the mean magnetic field modulus of
HD~75445 suggested by \citetalias{2022MNRAS.514.3485G} is not observed when
using the revised values of $\Bm$ that we derive from re-analysis of
the spectra studied by \citetalias{2017A&A...601A..14M}. This is
illustrated in Fig.~\ref{fig:bm_hjd_75445}. In this figure, the $\Bm$
values determined by \citetalias{2022MNRAS.514.3485G} from our CES
spectrum of the present study and from ESO Archive HARPS and UVES
spectra are replaced by our own measurements of these data, for better
overall consistency. Comparing Fig.~\ref{fig:bm_hjd_75445} with
Fig.~15 of G22, the linear trend emphasised by these authors is no
longer apparent. The rotation period of HD~75445 must vastly
exceed the $\sim$13\,yr time interval spanned by the analysed
spectra, or the star must be seen rotation pole on.

\begin{figure}
  \centering
  \includegraphics[width=1.0\linewidth,angle=0]{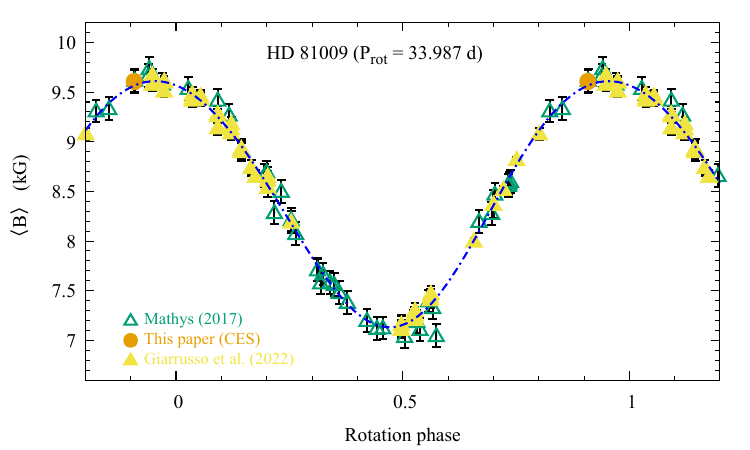}
\caption{Mean magnetic field modulus of HD~81009
  against rotation phase 
  computed with the value $\Prot=33\fd987$ of the rotation period
  \citep{2017PASP..129j4203P}. The phase origin has been
  set at JD~2\,448\,646.878, which is close to a minimum in the
  Str\"omgren $v$ band. The
  dash-dotted blue curve represents the 
  least-squares fit of the measurements by a cosine wave and its first
  harmonic. Different  
  point types and colours are used to distinguish the observations
  from different origins.}
\label{fig:bm_phi_81009}
\end{figure}

\subsection{HD~81009 (HR~3724; A3p CrSrSi)}
\label{sec:hd_81009}
The new data set analysed here only contains one additional CES
spectrum of HD~81009. However, this is one of the stars for which
the reanalysis of 
\citetalias{2017A&A...601A..14M} yields the most significant improvement in
the consistency of the derived $\Bm$ values. This field moment shows
variations with the rotation period determined by 
\citet{2017PASP..129j4203P} (see Fig.~\ref{fig:bm_phi_81009}). The
best fit of the observed $\Bm$ variations includes a small but 
formally significant (at the $6.3\,\sigma$ level) contribution of the
first harmonic.

\begin{figure}
  \centering
  \includegraphics[width=1.0\linewidth,angle=0]{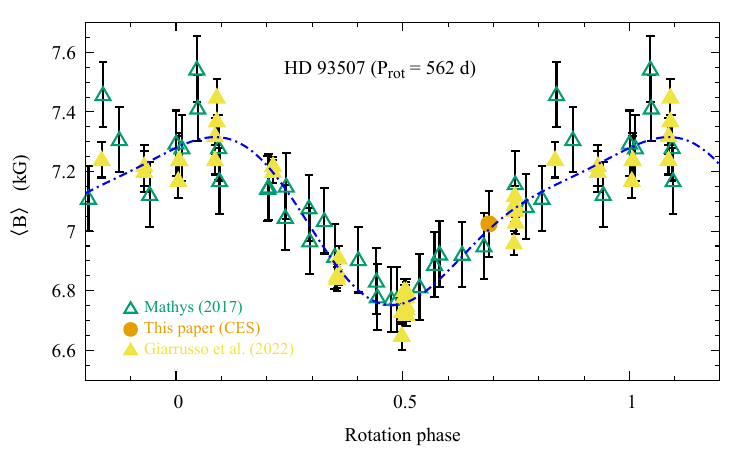}
\caption{Mean magnetic field modulus of HD~93507
  against rotation phase 
  computed with the value $\Prot=562$\,d of the rotation period
  \citepalias{2022MNRAS.514.3485G}. The phase origin has been
  set at JD~2\,448\,965.0, which is close to a $\Bm$ maximum. The
  dash-dotted blue curve represents the 
  least-squares fit of the measurements by a cosine wave and its first
  harmonic. Different  
  point types and colours are used to distinguish the observations
  from different origins.}
\label{fig:bm_phi_93507}
\end{figure}

\subsection{HD~93507 (CPD$-$67~1494; A0p SiCr)}
\label{sec:hd_93507}
The addition of a single CES spectrum of HD~93507 to the dataset
analysed by \citetalias{2022MNRAS.514.3485G} does not affect the value of
the rotation period derived by these authors, $\Prot=562$\,d. However,
for the best fit of the $\Bm$ variation curve, we include both the
fundamental cosine wave and the first harmonic. The amplitude of the
latter, which was omitted by \citetalias{2022MNRAS.514.3485G}, is
$\sim$0.25 that of the former; it is significant at the 
4.8\,$\sigma$ level. This fit is illustrated in
Fig.~\ref{fig:bm_phi_93507}.

\begin{figure}
  \centering
  \includegraphics[width=1.0\linewidth,angle=0]{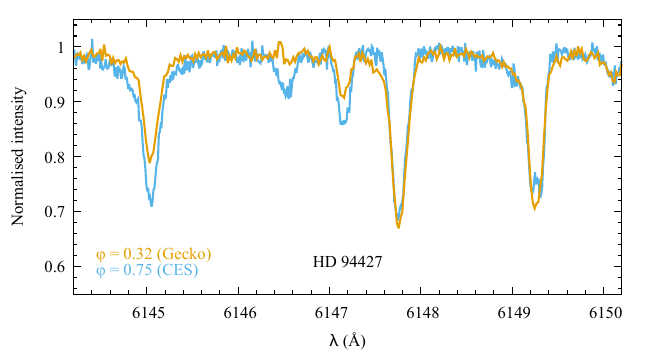}
  \caption{Portion of our CES spectrum (in light blue) and our Gecko
    spectrum (in orange) of
    HD~94427. The three main lines are Nd~{\sc
      iii}~$\lambda$\,6145.1\,\AA, Fe~{\sc ii}~$\lambda$\,6147.7\,\AA,
    and \Feline. Their Zeeman patterns are shown in Fig.~3 of
    \cite{2024A&A...691A.186M}. The phases at which these
    spectra were obtained were computed with the rotation period
    of \citet{2007A&A...475.1053A}, $\Prot=1\fd9625$, using the date
    of the first observation of these authors as phase origin
    (${\rm JD}_0=2452678.55$). While small, the resolution of the \Feline\
    doublet is clearly seen in the CES spectrum; it is only marginal
    in the Gecko spectrum, indicating that the mean magnetic field
    modulus is considerably weaker at the phase of the latter than at
    that of the former. The lines Cr~{\sc ii}~$\lambda\,6147.2$\,\AA\
    and La~{\sc ii}~$\lambda\,6146.5$\,\AA\ show remarkable
    variability. Both are stronger when $\Bm$ is stronger. In
    particular, the La~{\sc ii} line then becomes unusually strong,
    much more so than in most Ap stars.}
\label{fig:hd94427_6149}
\end{figure}

\subsection{HD~94427 (BD$-$11~2970; A5p SrEuCr)}
\label{sec:hd_94427}
We only have one CES spectrum and one Gecko spectrum of
HD~94427. Comparing them reveals the remarkable variability of this
star (see Fig.~\ref{fig:hd94427_6149}). Although small, the magnetic
splitting of the \Feline\, line is clearly resolved in the former,
allowing the value of the mean magnetic field modulus to be determined
with rather good precision. This resolution is reported here for the
first time. It is much more marginal in the Gecko spectrum, from
which the fit of the line profile by two similar gaussians still
allows the value of $\Bm$ to be estimated, but with a large
uncertainty. 

The magnetic field of HD~94427 had originally been detected by
\citet{2007A&A...475.1053A}, using spectropolarimetric
observations. They determined the rotation period of the star,
$\Prot=1\fd9625$, from eight measurements of its mean
longitudinal 
magnetic field. Comparing the CES spectrum in
Fig.~\ref{fig:hd94427_6149} (HD~94427; 
$\Bm=2.1$\,kG) with Fig.~\ref{fig:hd12098_6149} (HD~12098;
$\Bm=7.0$\,kG) illustrates how the lower rotational broadening in the
former allows the magnetic splitting of the \Feline\ line to be more
cleanly resolved than in the latter, which corresponds to a much
stronger field.  The (marginally) resolved line profile of this line
that we observe in HD~94427
can only be reconciled with the very short period
if the inclination angle $i$ of the rotation axis to the line of sight
is very small. This conclusion is consistent with the value of the
  inclination angle of the rotation axis to the line of sight derived
  by \citet{2007A&A...475.1053A}, $i=8\degr\pm4\degr$.

With $\Prot=1\fd9625$, the phase difference
between our two spectra of HD~94427 is close to half a cycle
($\Delta\phi=0.44$). With the large change in the value of $\Bm$ and
in the strengths of the Cr~{\sc ii}~$\lambda\,6147.2$\,\AA\
    and La~{\sc ii}~$\lambda\,6146.5$\,\AA\ lines, one might suspect
    that these observations were carried out close to the phases of
    the extrema of $\Bm$. However, these phases are close to those at
    which the mean longitudinal field reverses its sign, while in
    almost all Ap stars, the $\Bm$ and $\Bz$ extrema tend to be close
    to each other \citepalias[see Fig.~11
    of][]{2017A&A...601A..14M}. Significant departure from this trend
    would imply that the magnetic field of HD~94427 has a very unusual
    structure. Further observations are required to confirm this
    result and rule out the possibility of misinterpretation of the
    data.

\begin{figure}
  \centering
  \includegraphics[width=1.0\linewidth,angle=0]{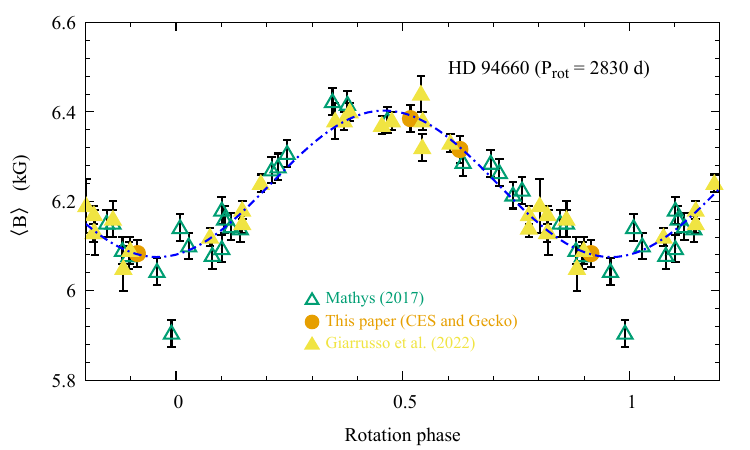}
\caption{Mean magnetic field modulus of HD~94660
  against rotation phase 
  computed with the value $\Prot=2830$\,d of the rotation period
  \citepalias{2022MNRAS.514.3485G}. The phase origin has been
  set at JD~2\,447\,000.0, which is close to a $\Bm$ minimum. The
  dash-dotted blue curve represents the 
  least-squares fit of the measurements by a cosine wave. Different 
  point types and colours are used to distinguish the observations
  from different origins.}
\label{fig:bm_phi_94660}
\end{figure}

\subsection{HD~94660 (HR 4263; A0p EuCrSi)}
\label{sec:hd_94660}
Combining our mean magnetic field measurements of HD~94660 with 22
values derived by \citetalias{2022MNRAS.514.3485G}, we confirm their value
of the rotation period, $\Prot=2830$\,d. Figure~\ref{fig:bm_phi_94660}
shows the corresponding 
phase diagram. The variation curve shows no significant deviation from
a sinusoid; the hint of anharmonicity reported by
\citetalias{2017A&A...601A..14M} is not confirmed.

\subsection{HD~110066 (HR~4816; A1p SrCrEu)}
\label{sec:hd_110066}
One of our two Gecko spectra of HD~110066 had already been analysed
by \citetalias{2022MNRAS.514.3485G}; we measured both of them
anyway. We also 
repeated the determination of the mean magnetic field modulus from the
four ESPaDOnS spectra studied by these authors, and added $\Bm$ values
derived from six more ESPaDOnS spectra from the CFHT
Archive. These new measurements and those of the spectra considered by
\citetalias{2017A&A...601A..14M} still do not indicate any variability of
the mean magnetic field modulus over a time span of $\sim$33\,yr. This
is consistent with the lack of 
variability of the mean longitudinal magnetic field
\citep{2014AstBu..69..427R,2017AstBu..72..391R,2020AstBu..75..294R,2022AstBu..77...94R}
and the absence of photometric variations greater than 0.01\,mag over
a 12\,yr time span \citep{2017PASP..129j4203P}. Either this star 
has an exceptionally long rotation period ($\Prot\gg33$\,yr), or it is
seen pole-on. 

\begin{figure}
  \centering
  \includegraphics[width=1.0\linewidth,angle=0]{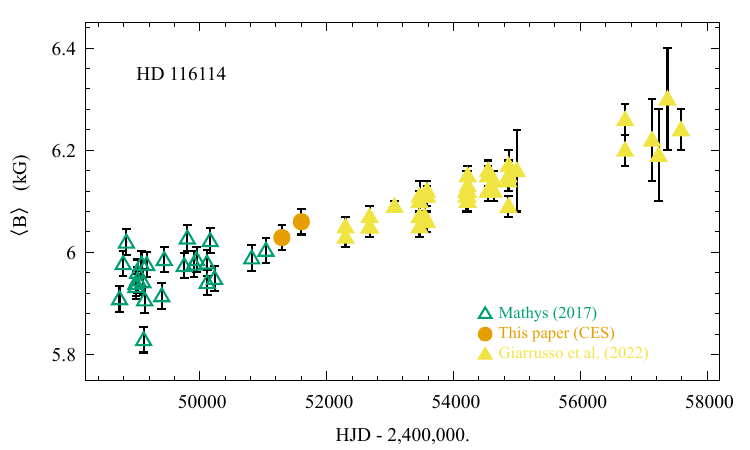}
\caption{Mean magnetic field modulus of HD~116114
  against Heliocentric Julian Date. Different
  point types and colours are used to distinguish the observations
  from different origins.}
\label{fig:bm_hjd_116114}
\end{figure}

\subsection{HD~116114 (BD$-$17~3829; F0p SrCrEu)}
\label{sec:hd_116114}
By combining mean magnetic field modulus measurements with those of
\citetalias{2017A&A...601A..14M}, \citetalias{2022MNRAS.514.3485G} established
that the rotation period of HD~116114 must be at least of the order of
48\,yr. The two additional $\Bm$ measurements that we derived from CES
spectra of 1999 April and 2000 February fit nicely in a gap of the
variation curve shown in their Fig.~20. This is illustrated here in
Fig.~\ref{fig:bm_hjd_116114}.

\subsection{HD~119027 (CD$-$28~10204; A3p SrEu)}
\label{sec:hd_119027}
Compared with the measurements of the 1993--1996 spectra of
\citetalias{2017A&A...601A..14M}, the analysis of the two additional
2000--2002 spectra hints at a small but significant monotonic
decrease of the values of both $\Bm$ and $\vr$, suggestive of
  both a rotation period and an orbital period considerably longer
  than 9 years. More spectra are currently being 
obtained. They will be analysed together with archive data to confirm
and characterise the variability of both quantities.

\begin{figure}
  \centering
  \includegraphics[width=1.0\linewidth,angle=0]{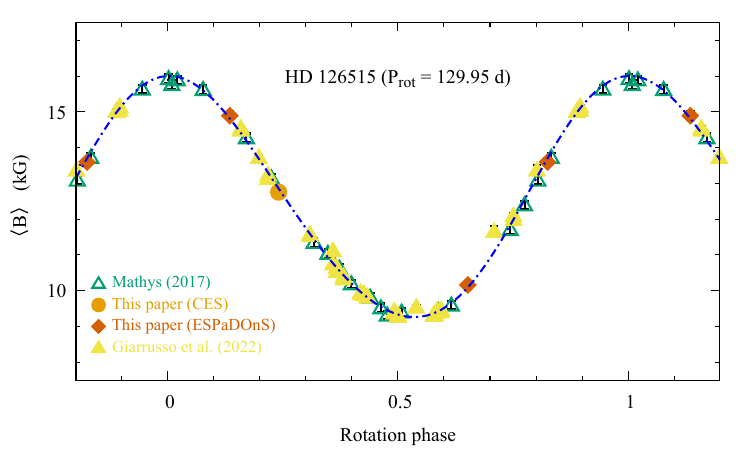}
\caption{Mean magnetic field modulus of HD~126515
  against rotation phase 
  computed with the value $\Prot=129\fd95$ of the rotation period
  \citepalias{1997A&AS..123..353M}. The phase origin has been
  set at JD~2\,437\,015, which is close to a $\Bm$ maximum. The
  dash-dotted blue curve represents the 
  least-squares fit of the measurements by a cosine wave and its first
  harmonic. Different 
  point types and colours are used to distinguish the observations
  from different origins.}
\label{fig:bm_phi_126515}
\end{figure}

\subsection{HD~126515 (BD+1~2927; A2p CrSrEu)}
\label{sec:hd_126515}
The measurements of the mean magnetic field modulus of HD~126515
reported by \citetalias{2022MNRAS.514.3485G} include one value derived
from the analysis of an ESPaDOnS archive spectrum. We repeated the
analysis of the latter, and added two more $\Bm$ determinations based
on ESPaDOnS archive spectra that became available more
recently. Considering also the $\Bm$ values based on the spectra
studied by \citetalias{2017A&A...601A..14M} and on an additional Gecko
spectrum, the mean field modulus variations are fully consistent with
the value of the rotation period confirmed by
\citetalias{2022MNRAS.514.3485G}, $\Prot=129\fd95$
\citepalias{1997A&AS..123..353M}. The corresponding phase diagram is shown
in Fig.~\ref{fig:bm_phi_126515}. 

\begin{figure}
  \centering
  \includegraphics[width=1.0\linewidth,angle=0]{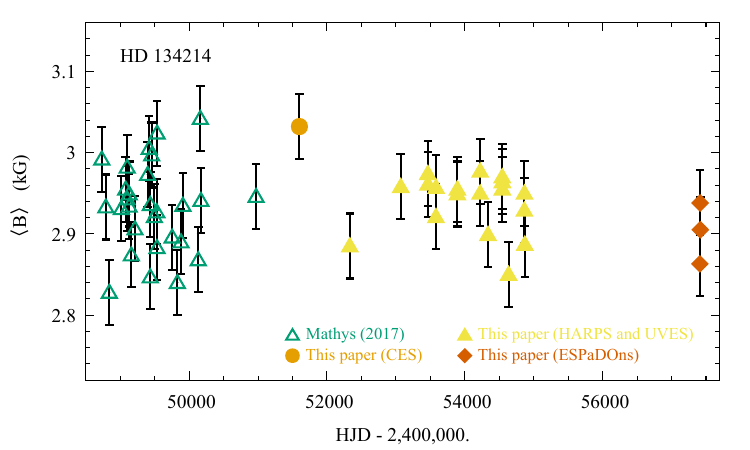}
\caption{Mean magnetic field modulus of HD~134214
  against Heliocentric Julian Date. Different
  point types and colours are used to distinguish the observations
  from different origins.}
\label{fig:bm_hjd_134214}
\end{figure}

\subsection{HD~134214 (BD$-$13~4081; F2p SrEuCr)}
\label{sec:hd_134214}
Figure~\ref{fig:bm_hjd_134214} shows 50 measurements of the mean
magnetic field modulus of HD~134214, from the spectra analysed by
\citetalias{2017A&A...601A..14M}, one additional CES spectrum, and HARPS,
UVES, and ESPaDOnS archive spectra. No obvious $\Bm$ variation is
seen. The mean longitudinal magnetic field measurements
published by Romanyuk's group
(\citealt{2014AstBu..69..427R}; \citealt{2020AstBu..75..294R,2022AstBu..77...94R}) 
are not indicative of any significant variation either. The rotation
period must be extremely long ($\Prot\gg23.5$\,yr) or the rotation
axis must be aligned with the line of sight.

\begin{figure}
  \centering
  \includegraphics[width=1.0\linewidth,angle=0]{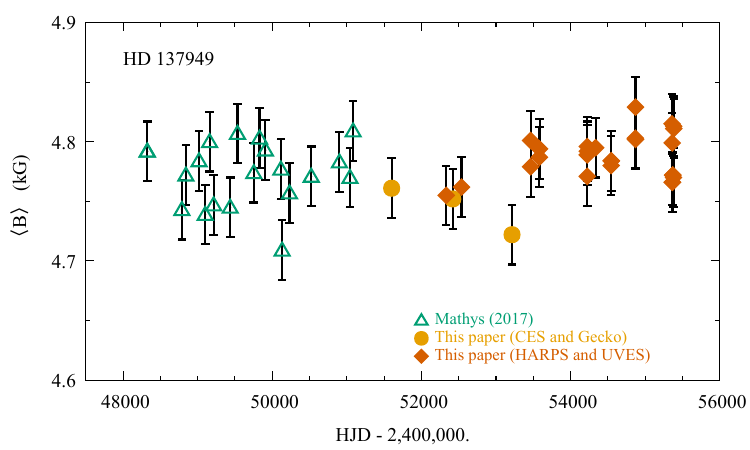}
\caption{Mean magnetic field modulus of HD~137949
  against Heliocentric Julian Date. Different
  point types and colours are used to distinguish the observations
  from different origins.}
\label{fig:bm_hjd_137949}
\end{figure}

\subsection{HD~137949 (BD$-$16~4093; F0p SrEuCr)}
\label{sec:hd_137949}
Given the remarkably small standard deviation of the mean magnetic
field modulus values over
7.5\,yr reported by \citetalias{2017A&A...601A..14M} (23\,G), we
remeasured the HARPS and UVES archive spectra analysed by
\citetalias{2022MNRAS.514.3485G} to ensure the best possible
consistency on an extended time base of more than 19\,yr. Our
determinations yield $\Bm$ values exceeding those of
\citetalias{2022MNRAS.514.3485G} by $\sim$80\,G on average. Plotting
them together with our measurements of the spectra of 
\citetalias{2017A&A...601A..14M} and of 3 additional CES and Gecko
observations from 2000--2004 (see Fig.~\ref{fig:bm_hjd_137949}), they
indicate that the mean magnetic field modulus of HD~137949 has not
shown any significant variation in 19.3\,yr. The standard deviation of
49 $\Bm$ values spanning this time interval is only
24\,G. The time base covered by these measurements is shorter than
  that considered by \citetalias{2022MNRAS.514.3485G}, which includes
  spectra obtained with additional instruments to which we do not
  have access. We do not combine the $\Bm$ values derived from
  analysis of the latter by \citetalias{2022MNRAS.514.3485G}, given
  the inconsistency between their measurements of the HARPS and UVES
  spectra and ours. This is why we conservatively adopt 19\,yr as the
  lower limit of the value of the rotation period, which is
  considerably less than the 27\,yr value proposed by
  \citetalias{2022MNRAS.514.3485G}.

The tentative period value $\Prot=14.2$\,yr suggested by
  \citetalias{2017A&A...601A..14M} is also ruled out, as it was based
  on mean longitudinal magnetic field measurements. Indeed,
  \citet{2022MNRAS.516.2629J} could not confirm other authors'
previous claims of variability of this field moment. The possibility
that the star can be seen pole-on cannot be ruled out.

\begin{figure}
  \centering
  \includegraphics[width=1.0\linewidth,angle=0]{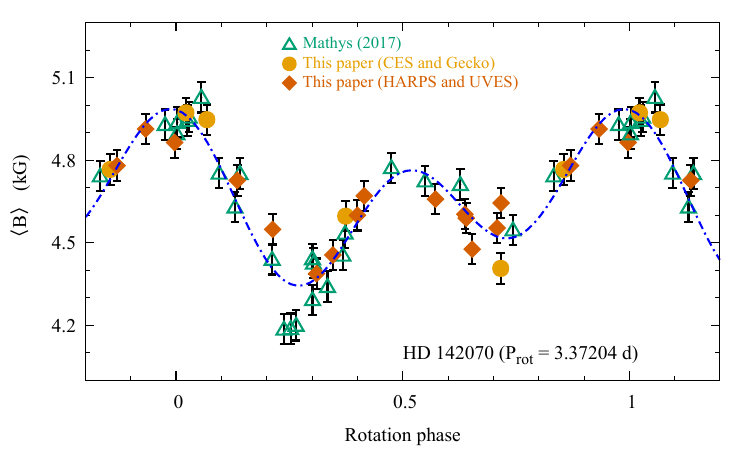}
\caption{Mean magnetic field modulus of HD~142070
  against rotation phase 
  computed with the value $\Prot=3\fd37204$ of the rotation period
  \citepalias{2022MNRAS.514.3485G}. The phase origin has been
  set at JD~2\,449\,878.2, which is close to a primary maximum of
  $\Bm$. The dash-dotted blue curve represents the 
  least-squares fit of the measurements by a cosine wave and its first
  harmonic. Different 
  point types and colours are used to distinguish the observations
  from different origins.}
\label{fig:bm_phi_142070}
\end{figure}

\subsection{HD~142070 (BD$-$0~3026; A0p SrCrEu)}
\label{sec:hd_142070}
While we originally intended to combine the mean magnetic field
modulus values
derived from the spectra from \citetalias{2017A&A...601A..14M} and those
determined here (from one CES spectrum and four Gecko spectra) with
the values published by 
\citetalias{2022MNRAS.514.3485G}, we realised that for the ESO Archive
spectra in common, the latter were partly inconsistent with those
that we obtained. The difference is particularly significant close to
the two minima of the $\Bm$ variation curve (see below), where it can
reach 300--400\,G. This suggests that it probably originates in the
difficulty of separating the two split components of the \Feline\
doublet because of their distortion by the rotational Doppler
effect. Indeed, the rotation period of HD~142070,
is exceptionally short for an Ap star with 
RMSL. The resulting $\Bm$ measurement discrepancy is not overly
worrisome, but in order to draw conclusions on the firmest possible
basis, we did not consider the values derived by
\citetalias{2022MNRAS.514.3485G}, to ensure that all $\Bm$ data in our
analysis were derived in a manner as uniform as possible. Thus, the
$\Bm$ phase diagram shown in Fig.~\ref{fig:bm_phi_142070} is based
exclusively on our measurements. The value of the period
  that we derive, $\Prot=3.37204\pm0.00006$\,d, only marginally
  differs from that of 
\citetalias{2022MNRAS.514.3485G}, $\Prot=3\fd3721$. 

\begin{figure}
  \centering
  \includegraphics[width=1.0\linewidth,angle=0]{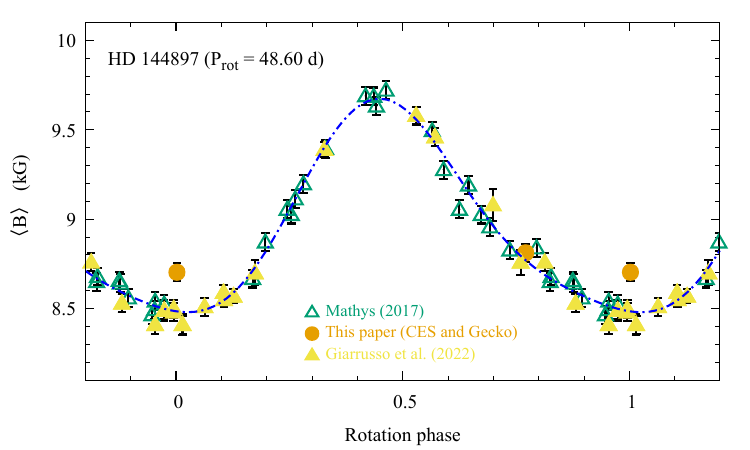}
\caption{Mean magnetic field modulus of HD~144897
  against rotation phase 
  computed with the value $\Prot=48\fd60$ of the rotation period
  \citepalias{2022MNRAS.514.3485G}. The phase origin has been
  set at JD~2\,447\,000.0, which is close to a $\Bm$ minimum. The
  dash-dotted blue curve represents the 
  least-squares fit of the measurements by a cosine wave and its first
  harmonic. Different 
  point types and colours are used to distinguish the observations
  from different origins.}
\label{fig:bm_phi_144897}
\end{figure}

\subsection{HD~144897 (CD$-$40~10236; B8p EuCr)}
\label{sec:hd_144897}
The combination of the 30 $\Bm$ values that we determined with the 
30 measurements of \citetalias{2022MNRAS.514.3485G} is fully consistent
with the value of the period derived by these authors,
$\Prot=48\fd60$. The corresponding variation curve is shown in
Fig.~\ref{fig:bm_phi_144897}.

\begin{figure}
  \centering
  \includegraphics[width=1.0\linewidth,angle=0]{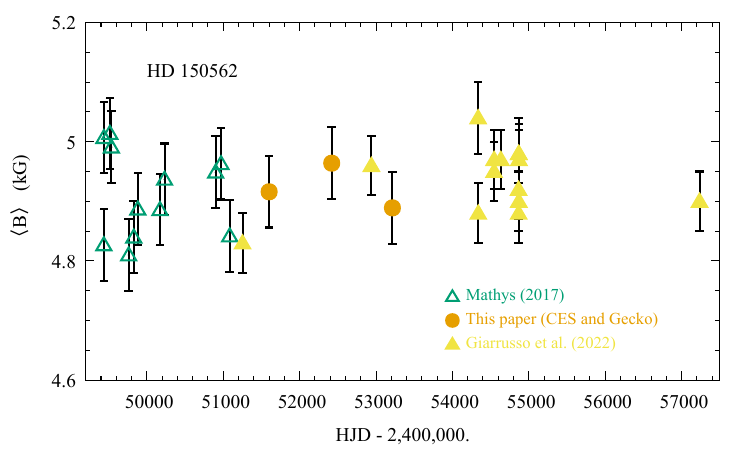}
\caption{Mean magnetic field modulus of HD~150562
  against Heliocentric Julian Date. Different
  point types and colours are used to distinguish the observations
  from different origins.}
\label{fig:bm_hjd_150562}
\end{figure}

\subsection{HD~150562 (CD$-$48~11127; A5:p EuSi?)}
\label{sec:hd_150562}
The 15 $\Bm$ values that we derived are plotted in
Fig.~\ref{fig:bm_hjd_150562} together with 
the 14 values determined by \citetalias{2022MNRAS.514.3485G}
(1999--2015). The behaviour apparent in this figure is not consistent
with the value of the rotation period, $\Prot=2100$\,d, proposed by
these authors. The periodogram built from the whole data set does not
show any clear 
peak. Figure~\ref{fig:bm_hjd_150562} rather suggests a slow monotonic
increase of $\Bm$ throughout the whole 21\,yr time spanned by the
observations, or at least over most of it. The impression of slight
decrease happening before the most recent observation must be
considered with caution, as it arises from a single point. The
rotation period of HD~150562 may plausibly exceed 21\,yr. 

\subsection{HD~157751 (CD$-$33~12069; B9p SiCr)}
\label{sec:hd_157751}
\citet{2007MNRAS.378L..16H} discovered 
RMSL in the spectrum of HD~157751, a star in which
\citet{2006AN....327..289H} had previously measured a significant mean
longitudinal magnetic field, $\Bz\sim4$\,kG. In our Gecko spectrum and
in a HARPS spectrum from the ESO-Archive, the \Feline\ line is only
marginally resolved, due to significant rotational
broadening. \citeauthor{2007MNRAS.378L..16H} estimated $\vsi=8$\,\kms,
which is consistent with the short rotation period derived by
\citet{2023ApJ...955..123S}, $\Prot=2\fd70134$, and a low
inclination of the rotation axis to the line of sight. The resulting
distortion of the spectral lines severely limits the achievable
precision in the determination of the mean magnetic field
modulus. However, the difference of $\sim$1\,kG between the value
derived by \citet{2007MNRAS.378L..16H}, $\Bm=6.6\pm0.5$\,kG, and our
measurements appears significant. As our two observations are
separated in phase by $\sim$0.5~cycle, and the spectrum analysed by
\citeauthor{2007MNRAS.378L..16H} was acquired at an intermediate phase,
it seems plausible that the latter may be in the vicinity of the $\Bm$
maximum.

\begin{figure}
  \centering
  \includegraphics[width=1.0\linewidth,angle=0]{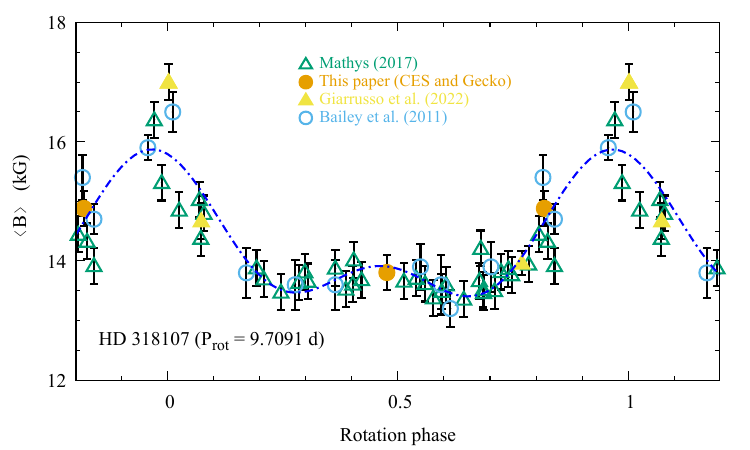}
\caption{Mean magnetic field modulus of HD~318107
  against rotation phase 
  computed with the value $\Prot=9\fd7091$ of the rotation period
  derived here. The phase origin has been
  set at JD~2\,448\,800.0, which is close to a primary $\Bm$
  maximum. The dash-dotted blue curve represents the 
  least-squares fit of the measurements by a cosine wave and its first
  harmonic. Different 
  point types and colours are used to distinguish the observations
  from different origins.}
\label{fig:bm_phi_318107}
\end{figure}

\subsection{HD~318107 (CD$-$32~13074; B8p)}
\label{sec:hd_318107}
One CES spectrum and one Gecko spectrum of HD~318107 were acquired
between 1999 and 2004. The latter was already studied by
\citetalias{2022MNRAS.514.3485G}; we remeasured it. Our analysis,
which also includes the $\Bm$ values determined by
\citet{2011A&A...535A..25B}, favours a value of the rotation period
$\Prot=9.7091\pm0.0001$\,d. This is only marginally longer than the value
adopted by \citetalias{2022MNRAS.514.3485G},
$\Prot=9.7089\pm0.0002$\,d. The variation of the mean magnetic field
modulus with the $\Prot=9\fd7091$ period is illustrated in
Fig.~\ref{fig:bm_phi_318107}. 

\begin{figure}
  \centering
  \includegraphics[width=1.0\linewidth,angle=0]{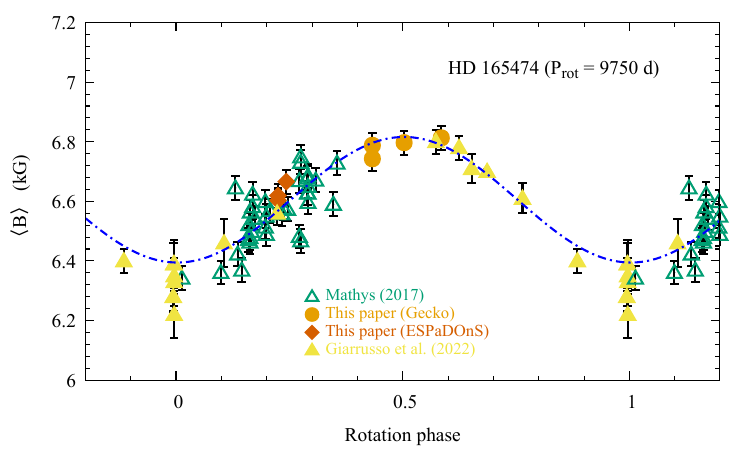}
\caption{Mean magnetic field modulus of HD~165474
  against rotation phase 
  computed with the value $\Prot=9750$\,d of the rotation period
  derived here. The phase origin, at JD~2\,447\,515.0, coincides with
  a $\Bm$ minimum. The dash-dotted blue curve represents the
  least-squares fit of the measurements by a cosine wave. Different
  point types and colours are used to distinguish the observations
  from different origins.}
\label{fig:bm_phi_165474}
\end{figure}

\subsection{HD~165474 (HR~6758B; A7p SrCrEu)}
\label{sec:hd_165474}
\citetalias{2022MNRAS.514.3485G} combined 37 mean magnetic field modulus
measurements of \citetalias{2017A&A...601A..14M} with 14 determinations
based on their analysis of spectra
of various origins to set a lower
limit $\Prot\gtrsim9900$\,d\ to the rotation period of HD~165474. Here
we complement this dataset with four $\Bm$ values derived from Gecko
spectra and three $\Bm$ values from
ESPaDOnS archive spectra. 

The time span covered by the
full data set, 33 years, is sufficient to determine an
accurate value 
of the rotation period, $\Prot=9750\pm125$\,d. 
Figure~\ref{fig:bm_phi_165474} shows the $\Bm$ variation curve with
this period. The analysed measurements cover
$\sim$ 1.2 
rotation cycle. The uncertainty of the period value was estimated by
plotting similar phase diagrams with series of progressively longer
and shorter periods, to constrain the range within which the
most recent $\Bm$ values remain consistent with the previous ones
under the assumption of the smooth magnetic variability typical of the
Ap stars. The addition four of new $\Bm$ values from Gecko
  spectra, which sample the phase range around the maximum, and of
  three new values from ESPaDOnS spectra, which improve the definition
  of the ascending branch of the variation curve, allow the period to be
  unambiguously established. That it is marginally shorter than the lower
  limit derived by \citetalias{2022MNRAS.514.3485G}, can be accounted
  for by the better constrained shape of the variation curve that we
  can achieve. 

At 26.7\,yr, the rotation period of HD~165474 is the second
  longest Ap star period with an accurate value, after HD~50169
($\Prot=29.04$\,yr; \citealt{2019A&A...624A..32M}). This
should probably be assigned to the time elapsed since
systematic studies of Ap stars with RMSL
have been started \citepalias{1997A&AS..123..353M}, a little more than 30
years ago. As more observations of other stars will be obtained in the
future, it will become possible to achieve accurate determinations of
longer periods.

\begin{figure}
  \centering
  \includegraphics[width=1.0\linewidth,angle=0]{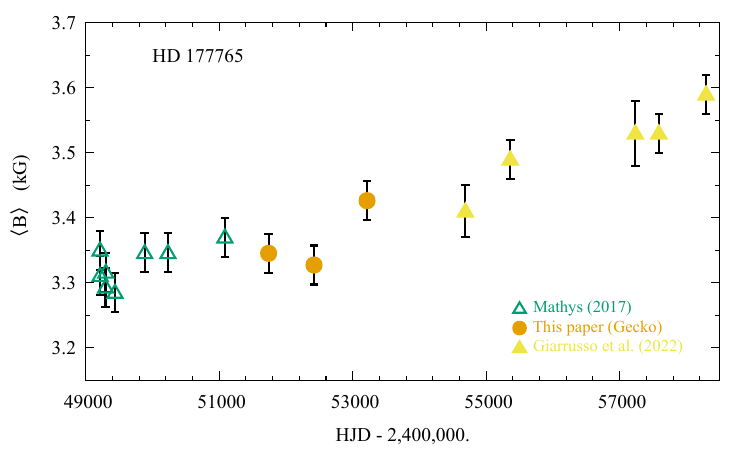}
\caption{Mean magnetic field modulus of HD~177765
  against Heliocentric Julian Date. Different
  point types and colours are used to distinguish the observations
  from different origins.}
\label{fig:bm_hjd_177765}
\end{figure}

\subsection{HD~177765 (CD$-$26~13816; A5p SrEuCr)}
\label{sec:hd_177765}
Combining our measurements of eight CES spectra and three more recent
Gecko spectra  
with five $\Bm$ determinations from \citepalias{2022MNRAS.514.3485G}, the
minimum of the mean magnetic field modulus of HD~177765 that seemed to
be visible in Fig.~31 of this reference is no longer
apparent. Instead, $\Bm$ 
seems to have been increasing monotonically since the discovery of
RMSL until the most recent
observation of this star, that is, over $\sim$25 years (see
Fig.~\ref{fig:bm_hjd_177765}). Thus, the rotation period must be
several decades long: under the assumptions that the variation curve
is symmetric about
its extrema and that the first and last measurements reported here are
close to, respectively, a minimum and a maximum of $\Bm$, 
the rotation period must be at 
least of the order of $\Prot\sim50$\,yr. However,
a much longer period is not implausible.

\begin{figure}
  \centering
  \includegraphics[width=1.0\linewidth,angle=0]{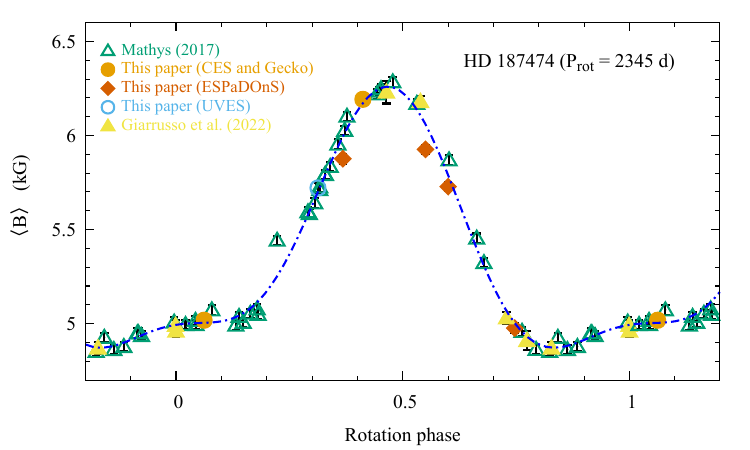}
\caption{Mean magnetic field modulus of HD~187474
  against rotation phase 
  computed with the value $\Prot=2345$\,d of the rotation period
  derived here. The phase origin has been
  set at JD~2\,446\,766. The
  dash-dotted blue curve represents the 
  least-squares fit of the measurements by a cosine wave and its first
  harmonic. Different 
  point types and colours are used to distinguish the observations
  from different origins.}
\label{fig:bm_phi_187474}
\end{figure}

\subsection{HD~187474 (HR~7552; A0p EuCrSi)}
\label{sec:hd_187474}
We combined mean magnetic field modulus values determined from the 40
high-resolution spectra analysed by \citetalias{2017A&A...601A..14M},
previously unpublished $\Bm$ determinations from one CES and one Gecko
spectrum, seven measurements from \citetalias{2022MNRAS.514.3485G}, and 
new determinations from \red {one UVES and four ESPaDOnS spectra} from the
respective archives. The resulting data set spans a time interval of
$\sim$35.5\,yr, from 1988 May to 2023 October. The value of the
rotation period derived from its analysis, $\Prot=2345\pm10$\,d, is
similar to the value $\Prot=2345\pm15$\,d determined by
\citet{1991A&AS...89..121M} from 50 mean longitudinal
magnetic field values spanning $\sim31$\,yr (1957 June to 1988
April). The uncertainty of our determination is considerably less
  than that of the value derived by \citetalias{2022MNRAS.514.3485G},
  $\Prot=2324\pm40$\,d. The $\Bm$ variation curve 
corresponding to $\Prot=2345$\,d is shown in
Fig.~\ref{fig:bm_phi_187474}. 

\subsection{HD~192678 (BD+53~2368; A2p Cr)}
\label{sec:hd_192678}
We have only one additional Gecko spectrum for HD~192678. This
spectrum was already analysed by \citetalias{2022MNRAS.514.3485G}. The
value that they determined for the mean magnetic field modulus does
not significantly differ from the one that we obtain. Accordingly, we
concur with their conclusions for this star.

\begin{figure}
  \centering
  \includegraphics[width=1.0\linewidth,angle=0]{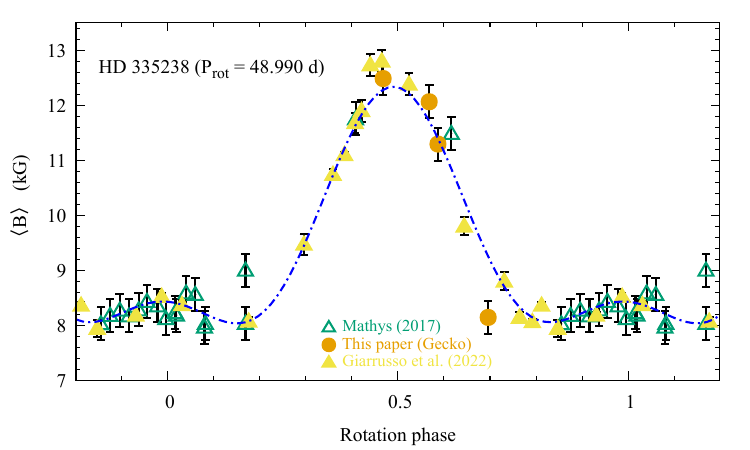}
\caption{Mean magnetic field modulus of HD~335238
  against rotation phase 
  computed with the value $\Prot=48\fd990$ of the rotation period
  derived here. The phase origin has been
  set at JD~2\,447\,450.0 so that the $\Bm$ maximum occurs near phase
  $\phi=0.5$. The dash-dotted blue curve represents the
  least-squares fit of the measurements by a cosine wave and its first
  harmonic. Different 
  point types and colours are used to distinguish the observations
  from different origins.}
\label{fig:bm_phi_335238}
\end{figure}

\subsection{HD~335238 (BD+29~4202; A1p CrEu)}
\label{sec:hd335238}
\citetalias{2022MNRAS.514.3485G} included the Gecko spectra that we
obtained in 2000 July and 2002 May in their study of HD~335238. We
carried out our own determinations of the mean magnetic field modulus
from these observations and from an additional Gecko
spectrum. Combining these data with the $\Bm$ values derived from the 
spectra considered by \citetalias{2017A&A...601A..14M} and the remaining
measurements of \citetalias{2022MNRAS.514.3485G}, the derived period 
of
$\Prot=48.990\pm0.010$\,d does not significantly differ from the
value adopted by \citetalias{2022MNRAS.514.3485G}
($\Prot=48.985\pm0.007$\,d). The corresponding phase diagram is shown
in Fig.~\ref{fig:bm_phi_335238}. The nearly perfect symmetry of the
variation curve about the phases of its extrema indicates that the
magnetic field of HD 335238 is axisymmetric. However, the large value
of the ratio between the primary $\Bm$ maximum and either of the two
$\Bm$ minima, $q\sim1.5$, shows that its structure must depart
considerably from a centred dipole.

\begin{figure}
  \centering
  \includegraphics[width=1.0\linewidth,angle=0]{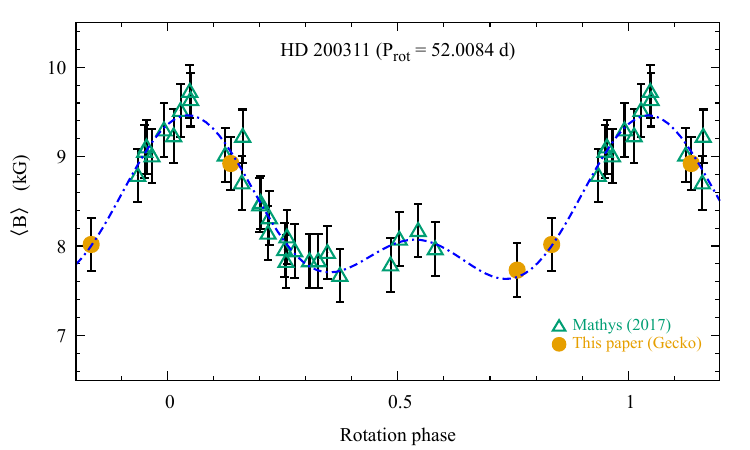}
\caption{Mean magnetic field modulus of HD~200311
  against rotation phase 
  computed with the value $\Prot=52\fd0084$ of the rotation period
  derived by \citet{1997MNRAS.292..748W}. The phase origin has been
  set at JD~2\,445\,407.513. The dash-dotted blue curve represents the
  least-squares fit of the measurements by a cosine wave and its first
  harmonic. Different 
  point types and colours are used to distinguish the observations
  from different origins.}
\label{fig:bm_phi_200311}
\end{figure}

\subsection{HD~200311 (BD+43~3786; B9p SiCrHg)}
\label{sec:hd_200311}
With the $\Prot=52.0084\pm0.0007$\,d rotation period determined by
\citet{1997MNRAS.292..748W} 
from the combination of photometry and mean longitudinal magnetic
field determinations, our three new
measurements of the mean magnetic field modulus of HD~200311 
nicely fill the main gap in the phase coverage of the data analysed by
\citetalias{2017A&A...601A..14M} (see Fig.~\ref{fig:bm_phi_200311}). The
addition of these measurements significantly improves the definition
of the variation curve. The latter draws a nearly symmetric double
wave. While not unexpected \citepalias[see][]{1997A&AS..123..353M},
this is seldom observed in Ap stars 
\citepalias{2017A&A...601A..14M}.

\begin{figure}
  \centering
  \includegraphics[width=1.0\linewidth,angle=0]{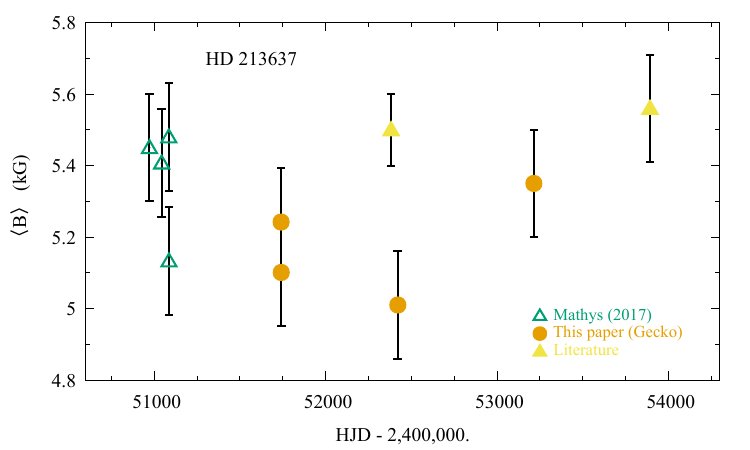}
\caption{Mean magnetic field modulus of HD~213637
  against Heliocentric Julian Date. Different
  point types and colours are used to distinguish the observations
  from different origins.}
\label{fig:bm_hjd_213637}
\end{figure}

\subsection{HD~213637 (BD$-$20~6447; F1p EuSr)}
\label{sec:hd_213637}
Our eight measurements of the mean magnetic field modulus of HD~213637
are plotted in Fig.~\ref{fig:bm_hjd_213637}, together with two
published measurements of UVES spectra recorded in 2002 April
\citep{2003A&A...404..669K} and 2006 June
\citep{2015MNRAS.446.4126E}. The former appears discrepant with our
data; the latter may be indicative of an actual field strength
increase. If so, the mean magnetic field modulus may have gone through
a minimum in 2002 and the rotation period may vastly exceed 8
yr. However, this is highly uncertain and needs to be confirmed. A
search for shorter periods does not yield any significant result.

\begin{figure}
  \centering
  \includegraphics[width=1.0\linewidth,angle=0]{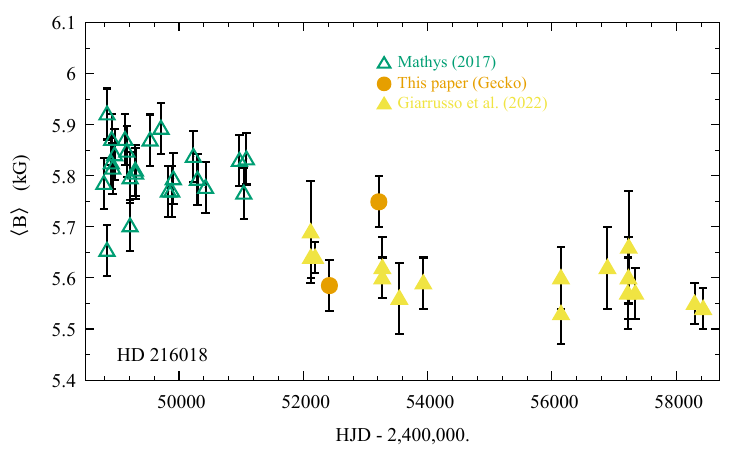}
\caption{Mean magnetic field modulus of HD~216018
  against Heliocentric Julian Date. Different
  point types and colours are used to distinguish the observations
  from different origins.}
\label{fig:bm_hjd_216018}
\end{figure}

\subsection{HD~216018 (BD$-$12~6357; A7p SrEuCr)}
\label{sec:hd_216018}
Our two Gecko spectra of HD~216018 from 2002 May and 2004 July had
already been analysed by \citetalias{2022MNRAS.514.3485G}. We also
measured them. The values that we derive for the mean magnetic field
modulus do not significantly differ from those obtained by
\citetalias{2022MNRAS.514.3485G}. We combined them with the 16
remaining $\Bm$ measurements of these authors and with the 24 $\Bm$
values derived from the analysis of the spectra considered by
\citetalias{2017A&A...601A..14M}. The period search carried out using the
resulting data set does not provide any convincing evidence in support
of the value $\Prot=34\fd044$ of the rotation period proposed by
\citetalias{2022MNRAS.514.3485G}. There is no
significant signal in the periodogram close to the $34\fd044$ value,
as already reported by \citet{2024A&A...683A.227M}.
The most prominent nearby peak corresponds to a period of $47\fd76$;
however it is not significant. Instead, consideration 
of Fig.~\ref{fig:bm_hjd_216018} suggests that the mean magnetic field
modulus has undergone a slow monotonic decrease since the first
observations of \citetalias{2017A&A...601A..14M} (obtained in 1992),
or shortly after, 
until the most recent ones of \citetalias{2022MNRAS.514.3485G} (2018),
hence that the rotation period of HD~216018 may vastly exceed the time
interval spanned by these data ($\sim$26\,yr).

\section{Discussion and conclusion}
\label{sec:disc}
As announced in the introduction, the main purpose of the study
reported in this paper is to ensure the conservation of the
information contents of high-resolution, natural light spectra of Ap
stars showing RMSL acquired between 1999
and 2004 that, in their majority, are not available in observatory
archives and had never been analysed until now. Our work is focused
on the determination of the mean magnetic field modulus (the primary
goal of the observing programmes as part of which these data were
obtained) and of the RV. Besides measuring these quantities in the
1999--2004 spectra, we repeated their determination in similar spectra
previously obtained within the framework of the same generic project
whose original analysis was reported by \citetalias{1997A&AS..123..353M}
and \citetalias{2017A&A...601A..14M} and, for some stars, we derived
additional values from CFHT and ESO archive spectra. While
substantial, our batch of new measurements does not justify a
systematic revision of the statistical trends derived in the
exhaustive studies of \citetalias{2017A&A...601A..14M} and
\citetalias{2022MNRAS.514.3485G}, with which our results are
consistent but for a minor adjustment (see below). The measurements
presented here valuably contribute to 
the improved knowledge of individual stars in several different
respects: rotation period, magnetic field variability and structure,
binarity, orbital period, and orbital elements. The main results in
these areas are summarised below.

The presence of RMSL in the stars
HD~12098 and HD~94427 is reported here for the first time. While this
detection is somewhat marginal, it is sufficient to determine the
value of the mean magnetic field modulus, albeit with a rather large
uncertainty.

The rotation period of HD~165474, $\Prot=26.71$\,yr, was
accurately determined for the first time. This is the second longest
Ap star period with an accurate value, after HD~50169
($\Prot=29.04$\,yr; \citealt{2019A&A...624A..32M}).
The rotation period of HD 29578 was unambiguously determined to be 
$\Prot=10.82$\,yr. The $\Bm$ variation curves of HD~29578 and
HD~165474 were also characterised (by least-squares fits). 
Reliable lower limits of the (long) rotation
periods were established for HD~47103 ($\Prot\gg6.8$\,yr), HD~116114
($\Prot\ga48$\,yr), HD~119027 ($\Prot\gg9$\,yr), HD~150562
($\Prot>21$\,yr), HD~177765 ($\Prot\ga50$\,yr), and HD~216018
($\Prot>26$\,yr). These stars show definite long-term drifts of their
mean magnetic field modulus, from which the $\Prot$ lower limits were
derived. The shorter periods proposed for some of them by
\citetalias{2022MNRAS.514.3485G} are confidently ruled out. The
long-term $\Bm$ drift reported by these authors for HD~55719 is
confirmed. By contrast, for four stars, we concluded that no
significant variation of the mean magnetic field modulus has been
observed 
over time spans of years: HD~75445 (over 13\,yr), HD~110066 (over
33\,yr), HD~134214 (over 23.5\,yr), and HD~137949 (over 19\,yr). We
cannot at present confidently decide if these stars are observed
rotation pole-on (and may accordingly have short or rather short
periods without showing rotational Doppler broadening of their
spectral lines), or if their rotation periods vastly exceed the time
interval spanned by their observations obtained until
now. Statistically, some of them likely have extremely long rotation
periods, as rotation axes aligned with the line of sight should be
rare. Only additional observations will allow us to distinguish
between the two possibilities on an individual basis. Linear
  spectropolarimetry should prove especially useful as a discriminant,
  as it samples the components of the magnetic vector in the plane
  orthogonal to the line of sight. The variation of the polarisation
  angle could be used to constrain the stellar rotation period.

Among the longest period stars of this study, HD~47103 represents the
only exception known to date to the conclusion reached by
\citetalias{2017A&A...601A..14M}, that the average over a rotation cycle of
the mean magnetic field modulus never exceeds 7.5\,kG in stars with
rotation periods longer than 150\,d. This potential exception had
not been ruled out by \citetalias {2017A&A...601A..14M} but could 
  not be definitely identified; it is now
confirmed. It does not detract from the fact that the rate of
occurrence of very strong fields ($\Bm_{\rm av}>7.5$\,kG) is much
higher in stars with relatively short periods ($\Prot\la150$\,d) than
in extremely long period stars. 

We confirmed that HD~200311 is a spectroscopic binary, as suggested by
\citetalias{2017A&A...601A..14M}; its orbital period vastly exceeds
12.5\,yr. Based on our RV measurements, 
HD~110066, HD~119027, HD~157751, and HD~335238 also appear to belong
to binaries. Longer-term followup is needed for definitive confirmation
of these results. In total, 19 of the 34 stars of our sample are
definite or probable binaries. In this paper, we determined for the
first time the orbital elements for three of them: HD~116114,
HD~142070, and HD~144897. All three have orbital periods of the order
of a decade and their orbital eccentricities, $\sim$0.65, do not
significantly differ from each other. We also refined the orbital
elements of six other stars, revised the lower limit of the
orbital period of HD~165474 , from $\Porb>6$\,yr
\citepalias{2017A&A...601A..14M} to $\Porb>33$\,yr, and set a lower limit
to the orbital period of HD~200311, $\Porb\gg12.5$\,yr. Of the 14
spectroscopic binaries whose
orbital period, or a lower limit thereof, has been determined, 12 have
$\Porb\ga2$\,yr. The two exceptions, HD~55719 and HD~61468, are also
the only two binaries in our sample that have eccentricities lower
than 0.2. The large eccentricities of the stars with long orbital
periods are consistent with the behaviour reported by
\citet{2002A&A...394..151C} for Ap stars in general.

In summary, the new $\Bm$ and $\vr$ measurements presented in
this paper usefully complement the previously available ones. Their
consideration allowed us to derive for the first time accurate,
unambiguous values
of two rotation periods longer than 10 years, to set considerably
increased lower limits to rotation periods of decades, to improve the
characterisation of the $\Bm$ variation curve of several stars, to
identify four new probable binary systems, to derive the orbital
elements for three spectroscopic binaries for which this had not been
previously done and 
to refine the orbital elements of six more, and to obtain new
constraints on the orbital periods of two systems. Overall, the
results of our analysis are consistent with the statistical
conclusions previously reached about the properties of Ap stars with
RMSL. The only exception is the
identification of a very strongly magnetic star ($\Bm_{\rm
  av}\sim17$\,kG) with a very long rotation period
($\Prot\gg6.8$\,yr). However, the existence of this star does not
question the conclusion that the rate of occurrence of very strong
magnetic fields ($\Bm_{\rm av}\ga7.5$\,kG) is much lower in very
slowly rotating stars ($\Prot\ga150$\,d) than in shorter period ones
\citepalias{2017A&A...601A..14M}. More generally, the measurements
presented here have a considerable legacy value as they represent an
important reference for future, long-term studies of the most slowly
rotating Ap stars and, more broadly, the understanding of the physical
processes responsible for the extreme differentiation (by a factor of
up to $10^5$--$10^6$) of the rotation rates among Ap stars (see
\citealt{2025A&A...703A.102M} and references therein for details).

\begin{acknowledgements}
Based on observations collected at the European Southern Observatory
under ESO programmes 062.L-0867 and 064.L-0443; and on observations
  collected at the Canada-France-Hawaii Telescope (CFHT), which is
  operated by the National Research Council of Canada, the Institut
  National des Sciences de l'Univers of the Centre National de la
  Recherche Scientifique of France, and the University of Hawaii. The
operations at the Canada-France-Hawaii Telescope are conducted with
care and respect from the summit of Mauna Kea, which is a significant
cultural and historic site. This research has also used processed data
obtained from the ESO Science Archive Facility; and 
the facilities of the Canadian Astronomy Data
Centre operated by the National Research Council of Canada with the
support of the Canadian Space Agency. 
This research has made use of the SIMBAD
database, operated at CDS, Strasbourg, France. JDL acknowledges the
financial support of the Natural Sciences and Engineering Research
Council of Canada (NSERC), funding reference
No. 6377-2016. T.\,A.\,A. Sigut  gratefully acknowledges support
from the NSERC through the Discovery Grant program.

\end{acknowledgements}

\bibliographystyle{aa}
\bibliography{bmref}

\begin{appendix}
  \nolinenumbers

\section{Observations and analysis}
\label{sec:obs}
To acquire the `new' spectra analysed here, the CES was used in the
Very Long Camera configuration fed by the ESO 3.6\,m telescope
\citep{2005CESVLC}. The Gecko spectra acquired at the CFHT are similar
to those obtained by \citetalias{1997A&AS..123..353M}. In these spectra,
the wavelengths of the resolved magnetic components of the \Feline\
line were measured by fitting them with gaussians, as described by
\citetalias{1997A&AS..123..353M}. Up to four gaussians were used, to
untangle the contributions of the \Feline\ line from those of the
blending lines affecting its blue wing (Sm~{\sc
  ii}~$\uplambda\,6149.06$\,\AA\ and an unidentified line, probably
from a rare earth element). 

The Zeeman pattern
of the \Feline\ line is a simple doublet and the separation of its
components is proportional to the mean magnetic field modulus of the
star ($\Bm$, the line-intensity weighted average over the
visible stellar disk 
of the modulus of the magnetic vector). The value of this field moment
can be determined in a mostly model-independent and approximation-free
manner by application of the formula
\begin{equation}
  \lambda_{\rm r}-\lambda_{\rm b}=g\,\Delta\lambda_{\rm Z}\,\Bm,
\label{eq:Bm}
\end{equation}
where $\lambda_{\rm r}$ and $\lambda_{\rm b}$ are the wavelengths of
the red and blue line components, respectively; $g=2.70$ is the Land\'e
factor of the split level of the transition; $\Delta\lambda_{\rm
  Z}=k\,\lambda_0^2$, with $k=4.67 \times 10^{-13}$\,\AA$^{-1}$\,G$^{-1}$;
and $\lambda_0=6149.258$\,\AA\ is the nominal wavelength of the
transition. The wavelengths are expressed in Angstr\"oms and the
magnetic field in Gauss.

Following \citetalias{2017A&A...601A..14M}, the above-described component
measurements were also used to determine the stellar radial velocity (RV),
$\vr$. To this effect, the observed
wavelength $\lambda_I$ of the \Feline line was computed by
averaging the 
wavelengths of its blue and red split components as measured for the
determination of the mean magnetic field modulus, as follows:
\begin{equation}
\lambda_I=(W_{\lambda,{\rm b}}\,\lambda_{\rm b}+W_{\lambda,{\rm
    r}}\,\lambda_{\rm r})/(W_{\lambda,{\rm b}}+W_{\lambda,{\rm r}})\,.
\label{eq:rv6149}
\end{equation}
Then, the wavelength difference $\lambda_I-\lambda_0$
was converted to an RV value in the standard manner.
The notations $W_{\lambda,{\rm b}}$ and $W_{\lambda,{\rm r}}$ refer
to the equivalent widths of the measured parts of each line
component. Indeed, the values of those equivalent widths depend on the
limits between which the
multiple Gaussians were fitted to the observed profile. These limits
in turn depend on line blends possibly affecting the \Feline\ line.

It is not straightforward to estimate the uncertainties of
measurements based on a single line. For the mean magnetic field
modulus, the way to proceed was discussed
in detail in Sect.~6 of \citetalias{1997A&AS..123..353M}. We followed the 
same approach. For each star, a single value of the uncertainty
$\sigma_B$ was used for all the mean magnetic field modulus values
that we 
derived. The adopted $\sigma_B$ values are listed in Table~A.1,
available in electronic form at the CDS. Column 1 contains the HD
number of the star and Col.~2 the corresponding $\sigma_B$ value. For
the RV values, following \citetalias{2017A&A...601A..14M}
(Sect.~5.6), the uncertainty was uniformly set to 1\,\kms. As
  stated by \citet{2020A&A...636A...6M}, for a given star, the
  achievable precision 
  in the measurements of $\lambda_{\rm b}$ and $\lambda_{\rm r}$ is
  almost independent of the spectral resolution and of the S/N of the
  observations, as long as both are sufficiently high. This ensures
  consistency in the derived values of $\Bm$ and $\vr$ between
  observations obtained at different epochs with different
  instrumental configurations. However, exceptional occurrences of
  unrecognised corruption of the \Feline\ line profile at isolated
  epochs (for instance, by the random hit of a high-energy particle on
  some detector pixels)
  cannot be ruled out. They may account for odd anomalous
  measurements, such as seen at JD~2\,450\,171 in
  Fig.~\ref{fig:bm_hjd_75445} or at phase 0.990 in
  Fig.~\ref{fig:bm_phi_94660}.

To ensure the best consistency between the measurements of the
additional spectra studied here and those of the high-resolution,
unpolarised spectra analysed
by \citetalias{1997A&AS..123..353M} and \citetalias{2017A&A...601A..14M}, we
systematically remeasured all the latter, except for the observations
obtained in 1993 June and 1994 December at the Kitt Peak National
Observatory with 
the 0.9\,m coud\'e feed telescope and the coud\'e spectrograph, which
are missing from our records and were omitted from the present
analysis. The resulting $\vr$ values do not 
significantly differ from the original ones, but the differences
between the revised and old $\Bm$ values are not negligible in some
cases. 

Furthermore, for some stars of our sample, we carried out additional
$\Bm$ determinations from
Canadian Astronomy Data Centre (CADC) archive spectra
from the CFHT Echelle SpectroPolarimetric Device for the Observation
of Stars (ESPaDOnS), and from ESO Archive
reduced spectra from the UV-Visual Echelle Spectrograph (UVES) and
the High Accuracy Radial velocity Planet Searcher (HARPS). This
analysis was restricted to spectra that had not been already measured
by \citet{2022MNRAS.514.3485G}, except for HD~110066, HD~137949, and
HD~142070, for which we reanalysed these data (see
Sects.~\ref{sec:hd_110066}, \ref{sec:hd_137949}, and
\ref{sec:hd_142070}). 

The
values of the mean magnetic field modulus that we determined are
presented in Table~A.2, available in electronic form at the CDS. The
star HD number is given in Col.~1, the Heliocentric Julian Date of the
observation in Col. 2, and the value of $\Bm$ in Col.~3. A code in
Col.~4 identifies the source of the measurement: 1 for the spectra
analysed in \citetalias{1997A&AS..123..353M} and
\citetalias{2017A&A...601A..14M}, 2 for the CES and Gecko 
spectra of the present paper, 3 for UVES archive spectra, 4 for HARPS
archive spectra, and 5 for ESPaDOnS archive spectra. 

As already mentioned, we combined all our $\Bm$ measurements with
those published by \citetalias{2022MNRAS.514.3485G} (except where
otherwise noted in
Sect.~\ref{sec:meas}) so as to achieve the
most complete characterisation to date of the mean magnetic field
modulus of the studied stars and of their variations. For those stars
for which the rotation period was known or could be determined, we
computed fits of the derived $\Bm$ values against rotation phase with a
cosine wave:
\begin{equation}
\Bm(\phi)=M_0+M_1\,\cos[2\pi\,(\phi-\phi_{M_1})]\,, 
\label{eq:fit1}
\end{equation}
or with the superposition of a cosine and of its first harmonic,
\begin{eqnarray}
\Bm(\phi)=M_0&+&M_1\,\cos[2\pi\,(\phi-\phi_{M_1})]\nonumber\\
&+&M_2\,\cos[2\pi\,(2\phi-\phi_{M_2})]\,.
\label{eq:fit2}
\end{eqnarray}
The harmonic was included only when the $M_2$ coefficient is formally
significant. The fit coefficients are given in
Table~\ref{tab:mfit}. The rotation phase is 
calculated using the values of the phase origin, HJD$_0$, and of the
rotation period, $\Prot$, that appear in
Table~\ref{tab:stars}. The fits are weighted by the
inverse of the square of the uncertainties of the individual
measurements. These fits are only meant to provide a simple
  numerical characterisation of the observed variations, without
  implication on the actual geometric structure of the magnetic
  field. Accordingly, we did not attempt to include higher-order
  harmonics to improve their quality. This may occasionally lead to
  apparent discrepancies around certain rotation phases, such as seen
  in Fig.~\ref{fig:bm_phi_142070}.

We complemented our RV determinations based on the high-resolution
spectra from \citetalias{2017A&A...601A..14M} and on the new CES and
Gecko spectra from the present paper with new $\vr$ values
derived from CADC ESPaDOnS archive spectra and from ESO Archive
reduced UVES, HARPS and FEROS spectra. (FEROS stands for
Fiber-fed Extended Range Optical Spectrograph.) The $\vr$
values were determined from measurements of the single line \Feline\
in these archive spectra, in the same way as described above.
Accordingly, we also assigned an
uncertainty of 1\,\kms\ to all these measurements. In
addition, in our analysis of the RVs, except where
otherwise noted, we included the $\vr$
values derived by \citetalias{2017A&A...601A..14M} from the analysis of
Stokes $I$ and $V$ spectra recorded with the Cassegrain Echelle
Spectrograph (CASPEC) fed by the ESO 3.6\,m telescope, as well as,
when available, $\vr$ data from the literature. The respective sources
of the latter are identified for each star in the relevant subsections
of Sect.~\ref{sec:SB}. 

Our RV measurements are
presented in Table~A.3, available in electronic form at the CDS. The
star HD number is given in Col.~1, the Heliocentric Julian Date of the
observation in Col. 2, and the value of $\vr$ in Col.~3. A code in
Col.~4 identifies the source of the measurement: 1 for the spectra
analysed in \citetalias{2017A&A...601A..14M}, 2 for the CES and Gecko
spectra of the present paper, 3 to 6 for UVES, HARPS, ESPaDOnS, and
FEROS archive spectra, respectively.

For the spectroscopic binaries in our sample whose orbital periods are
known or could be determined, we computed orbital solutions using the
Li\`ege Orbital Solution 
Package\footnote{\texttt{https://www.stsci.edu/\textasciitilde{}hsana/losp.html}}
(LOSP). The derived elements are given in Table~\ref{tab:orbits}. The
analysis of the spectroscopic binaries of our sample is further
discussed in Appendix~\ref{sec:SB}.

\section{Radial velocity variations and spectroscopic binaries}
\label{sec:SB}
In this appendix, we describe our analysis of those stars of our
sample that show RV variations resulting from binarity. We also
briefly review the RV variability of the rest of the sample.  

\subsection{HD~29578}
\label{sec:SB:hd_29578}
\begin{figure}
  \centering
  \includegraphics[width=1.0\linewidth,angle=0]{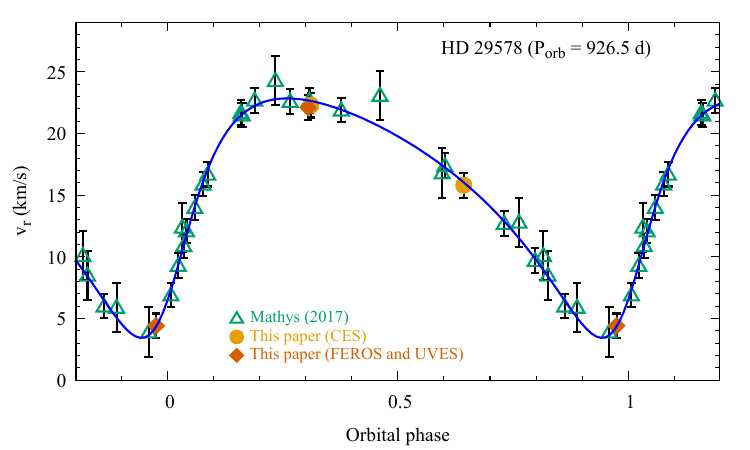}
\caption{Heliocentric radial velocity of HD~29578 against orbital 
  phase. The solid blue curve corresponds to the orbital solution
  given in Table~\ref{tab:orbits}. Different
  point types and colours are used to distinguish the observations
  from different origins.}
\label{fig:rv_phi_29578}
\end{figure}

For HD~29578, our RV data set includes four more values
than the one 
analysed by \citetalias{2017A&A...601A..14M}. This allowed us to refine
the determination of the orbital elements of this author. The updated
ones appear in Table~\ref{tab:orbits}, and the corresponding $\vr$
variability curve is shown in Fig.~\ref{fig:rv_phi_29578}.

\subsection{HD~47103}
\label{sec:SB:hd_47103}
The data set analysed here only contains one more RV
measurement of HD~47103 than the one studied by
\citetalias{2017A&A...601A..14M}. The new value is within the range spanned by
the previously existing data but its addition is insufficient to
constrain the orbit.

\subsection{HD~55719}
\label{sec:SB:hd_55719}
\begin{figure}
  \centering
  \includegraphics[width=1.0\linewidth,angle=0]{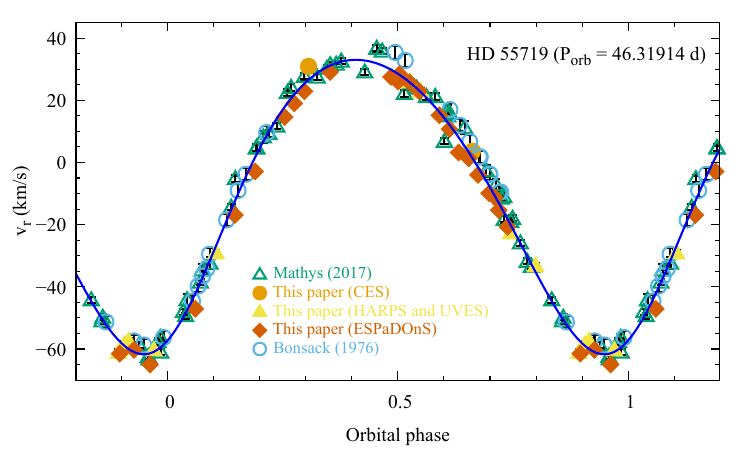}
\caption{Heliocentric radial velocity of HD~55719 against orbital 
  phase. The solid blue curve corresponds to the orbital solution
  given in Table~\ref{tab:orbits}. Different
  point types and colours are used to distinguish the observations
  from different origins.}
\label{fig:rv_phi_55719}
\end{figure}

New RV measurements of the primary Ap component of
HD~55719, a double-lined spectroscopic binary first reported by
\citet{1976ApJ...209..160B}, were carried out using the two
CES spectra mentioned in Sect.~\ref{sec:hd_55719} as well as ESO
Archive UVES and HARPS 
spectra and CFHT Archive ESPaDOnS spectra. This extends the time
interval spanned by available $\vr$ data to $\sim$42.9 years. The
RV curve corresponding to the revised orbital elements
that we computed from the full data set is shown in
Fig.~\ref{fig:rv_phi_55719}. An intriguing feature is the tendency for
the ESPaDOnS \vr\ values to lie systematically below the best-fit
curve. Given that these measurements, which have been obtained from
spectra recorded between 2013 and 2015, are considerably more recent
than all others, one may wonder if the systematic difference that they
seem to show may be indicative of the presence of a third, very
distant component in the HD~55719 system.

\subsection{HD~61468}
\label{sec:SB:hd_61468}
Inclusion of the new $\vr$ value that we derive from our additional
CES spectrum does not significantly improve the determination of the
orbital elements of the HD~61468 spectroscopic binary with respect to
those reported by \citetalias{2017A&A...601A..14M}. We present the revised
elements in Table~\ref{tab:orbits} for completeness. 

\subsection{HD~81009}
\label{sec:SB:hd_81009}
\begin{figure}
  \centering
  \includegraphics[width=1.0\linewidth,angle=0]{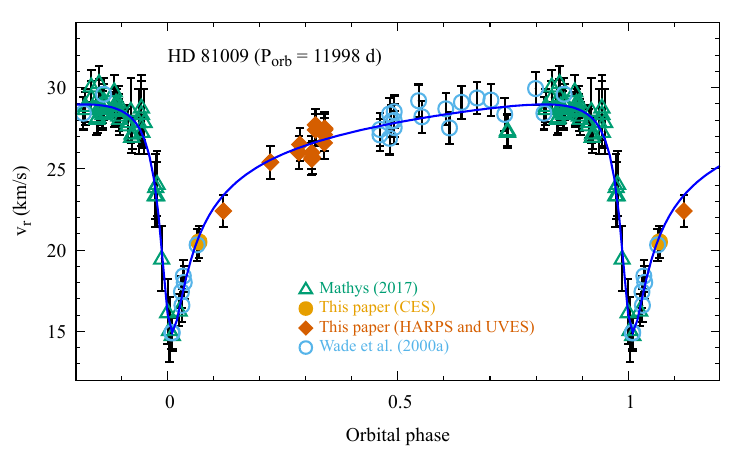}
\caption{Heliocentric radial velocity of HD~81009 against orbital 
  phase. The solid blue curve corresponds to the orbital solution
  given in Table~\ref{tab:orbits}. Different
  point types and colours are used to distinguish the observations
  from different origins.}
\label{fig:rv_phi_81009}
\end{figure}

\citet{2000A&A...361..991W} carried out the most recent attempt at
constraining the orbital elements of HD~81009 by combining astrometric
observations and RV measurements. They described in
detail the ambiguities that plagued for a long time the determination
of the orbital period. They solved them for the first time and
concluded unambiguously that this period length must be of the order
of 29.3\,yr. However, this value is considerably longer than the time
interval spanned by the RV data that were analysed. Here
we complement the previously existing $\vr$ measurements with new
ones from the CES spectrum mentioned in Sect.~\ref{sec:hd_81009} and
from UVES and HARPS 
reduced spectra from the ESO Archive, adding $\sim$9 years to the
overall time coverage. We carried out an updated determination of the
orbital elements using only the RV data, with the results
given in Table~\ref{tab:orbits} and illustrated in
Fig.~\ref{fig:rv_phi_81009}. The revised value of the orbital period
that we derived, $\Porb=(11998\pm618)$\,d, differs from the value
obtained by \cite{2000A&A...361..991W}, $\Porb=(10673\pm99)$\,d, by
more than can be accounted for by the formal errors. However,
considering 
that the existing RV measurements do not quite cover yet a full orbit,
both results consistently indicate a period length of the order of 30
years and definitely rule out other very different values that had
been previously proposed \citep[see][for
details]{2000A&A...361..991W}. 

\begin{figure}
  \centering
  \includegraphics[width=1.0\linewidth,angle=0]{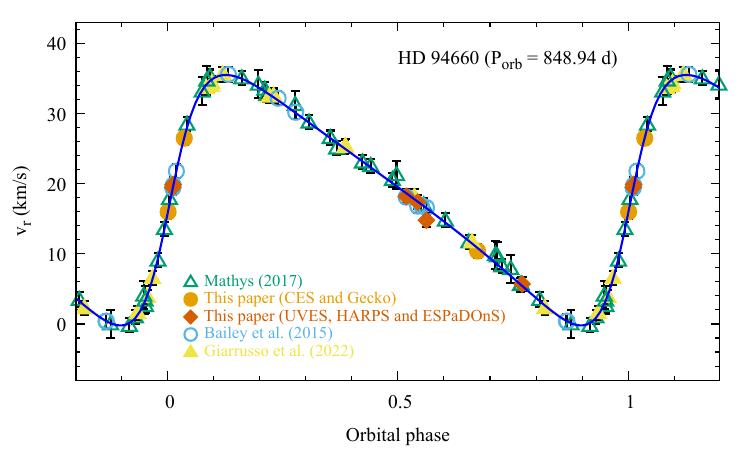}
\caption{Heliocentric radial velocity of HD~94660 against orbital 
  phase. The solid blue curve corresponds to the orbital solution
  given in Table~\ref{tab:orbits}. Different
  point types and colours are used to distinguish the observations
  from different origins.}
\label{fig:rv_phi_94660}
\end{figure}

\subsection{HD~94660}
\label{sec:SB:hd_94660}
For HD~94660, we analysed two HARPS spectra, three UVES spectra, and one
ESPaDOnS spectrum from the respective observatory archives to determine
$\vr$ at six additional epochs.
We combined the resulting data with our 26 RV
measurements from high-resolution spectra, 12
contemporaneous values obtained from CASPEC spectra
\citepalias{2017A&A...601A..14M}, 12 values derived
between 2001 May and 2014 April by \citet{2015A&A...575A.115B}, and 15
values from spectra obtained between 1992 December and 2009 November
and analysed by \citetalias{2022MNRAS.514.3485G}, to compute revised
orbital elements for HD~94660 (see Table~\ref{tab:orbits} and
Fig.~\ref{fig:rv_phi_94660}). The two new observations acquired in
2000 February and 2002 May, as well as the two HARPS spectra of 2009
May, correspond to two orbital phases between
0.0 and 0.05 at which the $\vr$ curve has the steepest slope, so that
they yield particularly stringent constraints on the value of the
orbital period, $\Porb=848.94\pm0.12$\,d. 

\begin{figure}
  \centering
  \includegraphics[width=1.0\linewidth,angle=0]{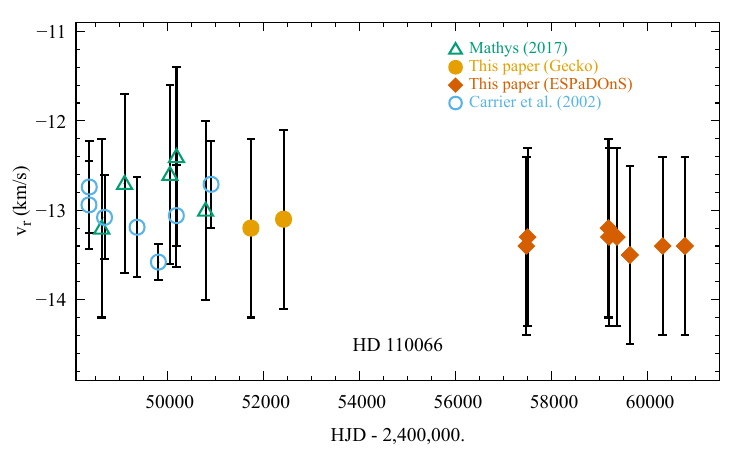}
\caption{Heliocentric radial velocity of HD~110066 against observation 
  date. Different
  point types and colours are used to distinguish the observations
  from different origins.}
\label{fig:rv_hjd_110066}
\end{figure}

\subsection{HD~110066}
\label{sec:SB:hd_110066}
Figure~\ref{fig:rv_hjd_110066} suggests that the RV of
HD~110066 may have become marginally more negative from the
group of measurements obtained between 1991 April and 2002 May
(including the spectra analysed by \citetalias{2017A&A...601A..14M} and
the $\vr$ values of \citealt{2002A&A...394..151C}) to
those of 2016 March--2025 April (based on ESPaDOnS spectra from the
CFHT Archive). The former average out at $-12.94$\,\kms\ with a
standard deviation of 0.32\,\kms; the latter at $-13.40$\,\kms\ with
a standard deviation of 0.09\,\kms. The significance of this
difference needs to be confirmed by additional long-term
observations. If the variability is real, it would be indicative of a
very wide binary, which might also lend itself to interferometric or
astrometric observations. We are not aware of the existence 
observations of this type for HD~110066.  

\begin{figure}
  \centering
  \includegraphics[width=1.0\linewidth,angle=0]{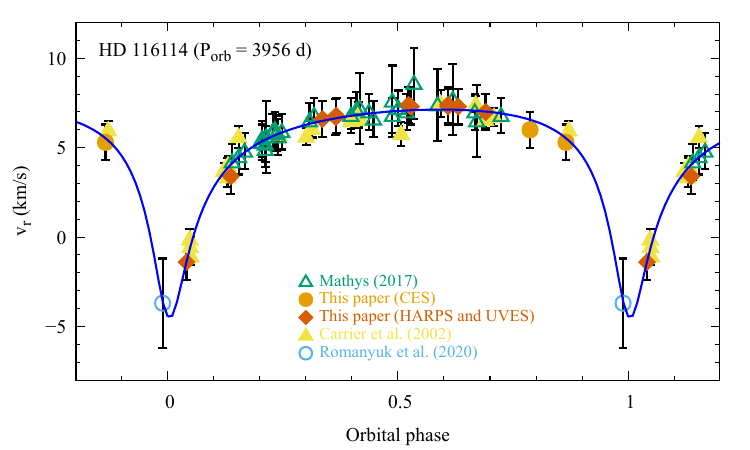}
\caption{Heliocentric radial velocity of HD~116114 against orbital 
  phase. The solid blue curve corresponds to the orbital solution
  given in Table~\ref{tab:orbits}. Different
  point types and colours are used to distinguish the observations
  from different origins.}
\label{fig:rv_phi_116114}
\end{figure}

\subsection{HD~116114}
\label{sec:SB:hd_116114}
\citetalias{2017A&A...601A..14M} had already reported that the observed
variations of the RV of HD~116114 are suggestive of a
binary with an orbital period of the order of 4000\,d. Here we combine
the $\vr$ data of his analysis \citep[including measurements
of][]{2002A&A...394..151C} with one more recent published value
\citep{2020AstBu..75..294R}, measurements from our two new CES
spectra, and measurements of ESO-Archive spectra obtained with UVES
(at two epochs) and with HARPS (at 13 epochs). Given the estimated
length of the orbital period, we used only one archive spectrum per
night even when more were available; in this case, the highest S/N
spectrum was preferred. Based on this data set, we could determine for
the first time the orbital elements of 
HD~116114 (see Table~\ref{tab:orbits}). The corresponding RV curve is 
shown in Fig.~\ref{fig:rv_phi_116114}. It is characterised by an
orbital period close to 11\,yr and a large eccentricity ($e=0.65$). 

\begin{figure}
  \centering
  \includegraphics[width=1.0\linewidth,angle=0]{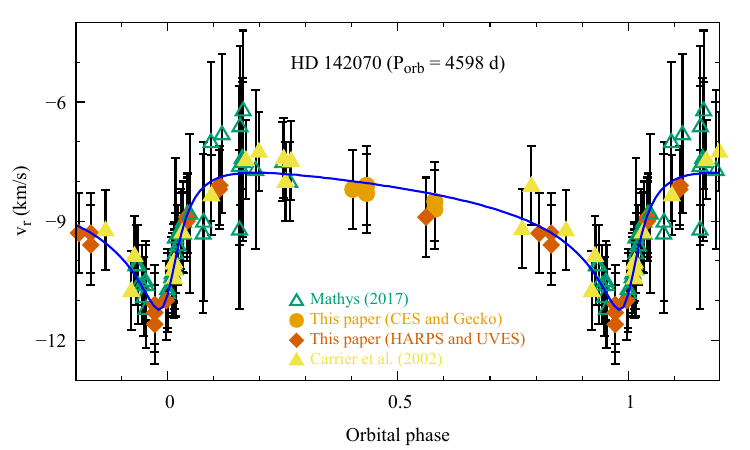}
\caption{Heliocentric RV of HD~142070 against orbital 
  phase. The solid blue curve corresponds to the orbital solution
  given in Table~\ref{tab:orbits}. Different
  point types and colours are used to distinguish the observations
  from different origins.}
\label{fig:rv_phi_142070}
\end{figure}

\subsection{HD~142070}
\label{sec:SB:hd_142070}
For HD~142070, new $\Bm$ and $\vr$ determinations were obtained from
one CES spectrum and four Gecko spectra. In addition, the RV
of the star was also measured at 15 epochs, using one
UVES spectrum and 14 HARPS spectra from the ESO Archive. (For nights
during which several spectra were obtained, we only used the one
recorded at the highest S/N, since denser sampling is not justified
given the length of the orbital period -- see below.)

Combining these new $\vr$ values with those derived from the spectra
analysed by \citetalias{2017A&A...601A..14M} and with the measurements of
\citet{2002A&A...394..151C}, we determined for the first time the
orbital elements of HD~142070. Figure~\ref{fig:rv_phi_142070} shows
the corresponding RV curve. The length of the orbital
period ($\Porb\sim12.6$\,yr) and the large eccentricity ($e=0.65$) must
be noted. 

\begin{figure}
  \centering
  \includegraphics[width=1.0\linewidth,angle=0]{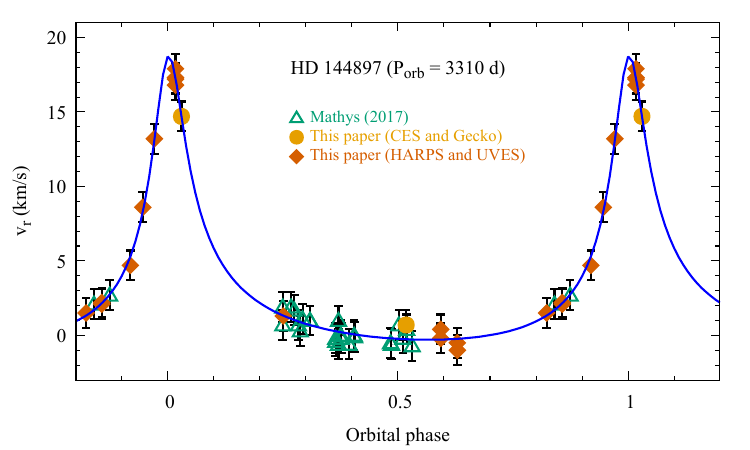}
\caption{Heliocentric radial velocity of HD~144897 against orbital 
  phase. The solid blue curve corresponds to the orbital solution
  given in Table~\ref{tab:orbits}. Different
  point types and colours are used to distinguish the observations
  from different origins.}
\label{fig:rv_phi_144897}
\end{figure}

\subsection{HD~144897}
\label{sec:SB:hd_144897}
We carried out RV determinations from our
`new' CES and Gecko spectra, as well as from one UVES
spectrum and 14 HARPS spectra retrieved from the ESO Archive. We
combined these measurements with the $\vr$ values derived from the
high-resolution spectra recorded in natural light from the study of
\citetalias{2017A&A...601A..14M}. (We did not include the CASPEC-based
measurements from this reference, as they needlessly increased the
scatter due to their more limited precision.) This allowed us to
determine, for the first time, the orbital elements of HD~144897 (see
Table~\ref{tab:orbits}). The result is illustrated in
Fig.~\ref{fig:rv_phi_144897}. The period length ($\Porb\sim9$\,yr) and
the high eccentricity ($e=0.67$) are noticeable. 

\begin{figure}
  \centering
  \includegraphics[width=1.0\linewidth,angle=0]{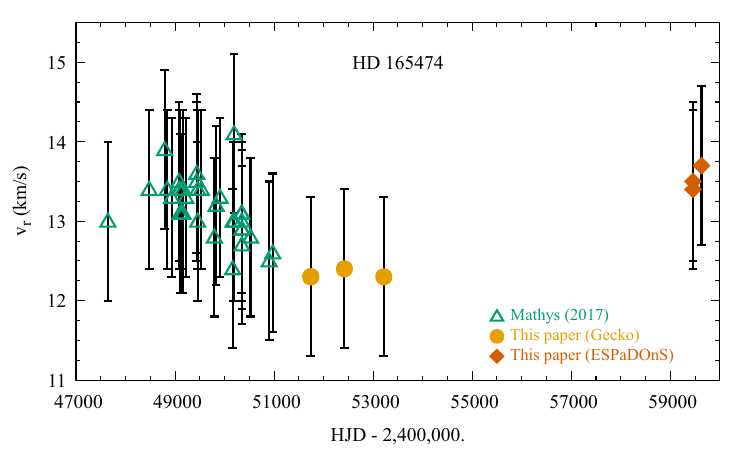}
\caption{Heliocentric radial velocity of HD~165474 against observation 
  date. Different
  point types and colours are used to distinguish the observations
  from different origins.}
\label{fig:rv_hjd_165474}
\end{figure}

\subsection{HD~165474}
\label{sec:SB:hd_165474}
\citetalias{2017A&A...601A..14M} emphasised the consistency of the RV
values derived as part of the $\Bm$ determination from
spectra of various origins and the resulting feasibility of detecting
minute variations even from such less than optimal single-line
measurements. The new measurements that we obtain here confirm these
conclusions. When plotted against observation date
(Fig.~\ref{fig:rv_hjd_165474}), they suggest that the orbital
period of HD~165474 may exceed 12\,000\,d. 

\begin{figure}
  \centering
  \includegraphics[width=1.0\linewidth,angle=0]{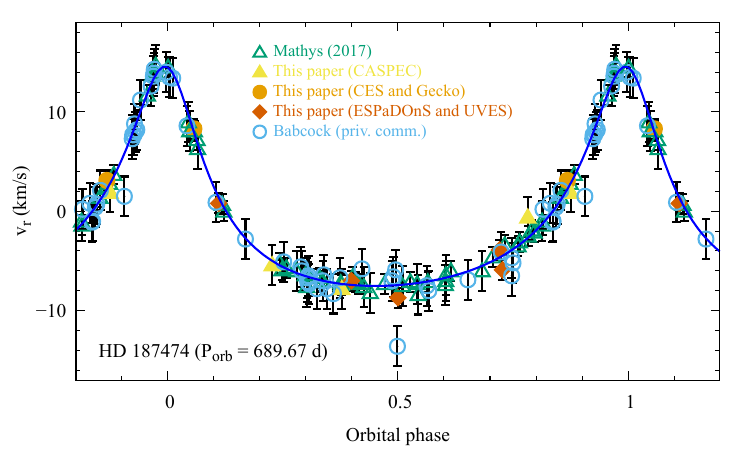}
\caption{Heliocentric radial velocity of HD~187474 against orbital 
  phase. The solid blue curve corresponds to the orbital solution
  given in Table~\ref{tab:orbits}. Different
  point types and colours are used to distinguish the observations
  from different origins.}
\label{fig:rv_phi_187474}
\end{figure}

\subsection{HD~187474}
\label{sec:SB:hd_187474}
We determined the RV of HD~187474 from 38 of the 40
high-resolution spectra
analysed by \citetalias{2017A&A...601A..14M}, one additional CES spectrum
and one additional Gecko spectrum, and one UVES Archive spectrum and four
ESPaDOnS Archive spectra. We complemented these determinations with
50 measurements from Babcock (private communication) and the
  20 CASPEC spectra considered by \citetalias{2017A&A...601A..14M}.
We used the resulting data set to compute
revised orbital elements (see Table~\ref{tab:orbits}). They represent a
small improvement with respect to those reported by
\citetalias{2017A&A...601A..14M}. The corresponding variability curve is
shown in Fig.~\ref{fig:rv_phi_187474}. 

\begin{figure}
  \centering
  \includegraphics[width=1.0\linewidth,angle=0]{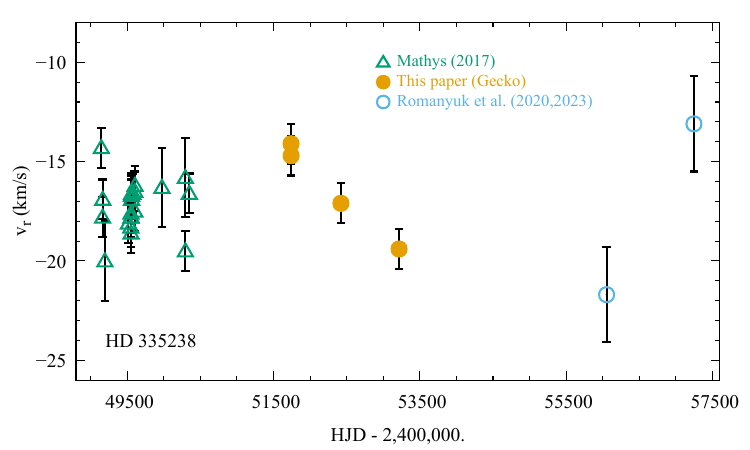}
\caption{Heliocentric radial velocity of HD~335238 against observation 
  date. Different
  point types and colours are used to distinguish the observations
  from different origins.}
\label{fig:rv_hjd_335238}
\end{figure}

\subsection{HD~335238}
\label{sec:SB:hd_335238}
While \citetalias{2017A&A...601A..14M} had not found any evidence of
variations of the RV of HD~335238, our new $\vr$
measurements of 2000--2004 support the detection of low-amplitude
variability
reported by \citet{2023AstBu..78..567R}. This is illustrated in
Fig.~\ref{fig:rv_hjd_335238}, where all the available RV
measurements (including one from \citealt{2020AstBu..75..294R}) are
plotted against observation date. Unfortunately, those data are
insufficient to derive accurate quantitative constraints about the
orbital elements, but a long period (hundreds to thousands of days)
seems plausible.

\begin{figure}
  \centering
  \includegraphics[width=1.0\linewidth,angle=0]{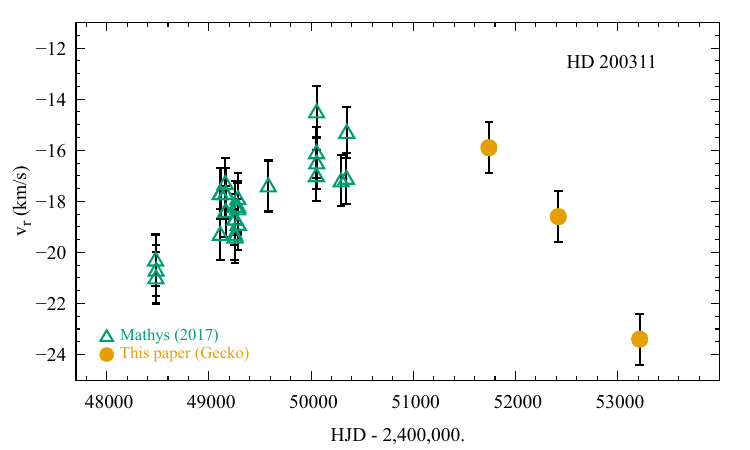}
\caption{Heliocentric radial velocity of HD~200311 against observation 
  date. Different
  point types and colours are used to distinguish the observations
  from different origins.}
\label{fig:rv_hjd_200311}
\end{figure}

\begin{figure}
  \centering
  \includegraphics[width=1.0\linewidth,angle=0]{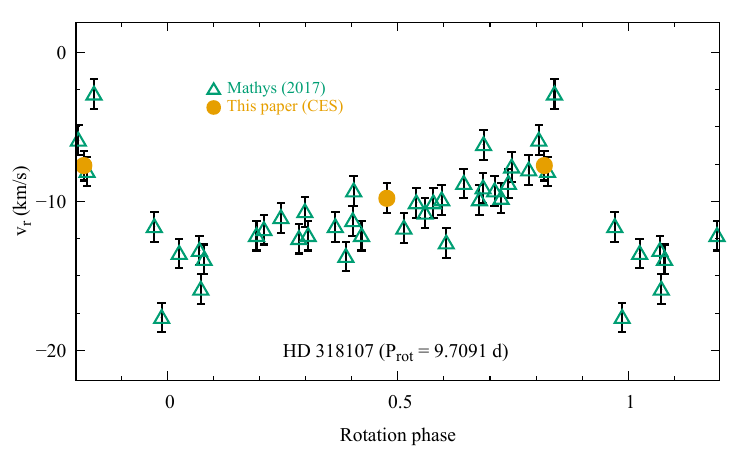}
  \includegraphics[width=1.0\linewidth,angle=0]{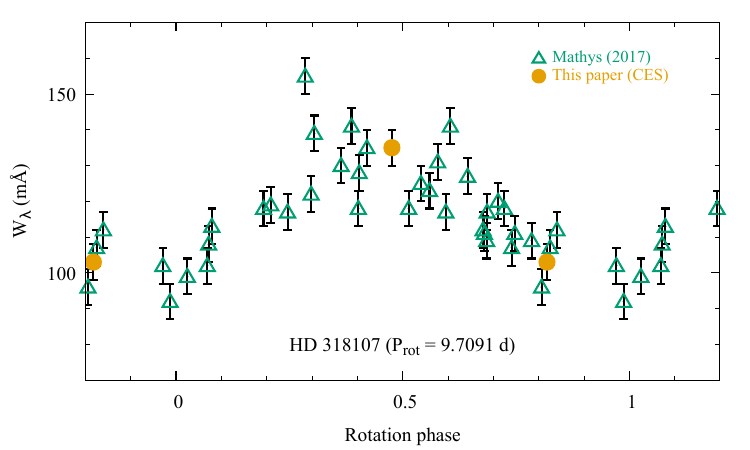}
\caption{Radial velocity derived from the \Feline\ line ({\em top\/})
  and equivalent width of the Fe~{\sc ii}\,$\lambda$\,6147.7\,\AA\
  line ({\em bottom\/}) in HD~318107 against the rotation phase
  computed with the same time origin and rotation period as in
  Fig.~\ref{fig:bm_phi_318107}. Different
  point types and colours are used to distinguish the observations
  from different origins.}
\label{fig:rv_ew_318107}
\end{figure}

\subsection{HD~200311}
\label{sec:SB:hd_200311}
Our new RV measurements fully confirm that HD~200311 is a
spectroscopic binary, as suggested by
\citetalias{2017A&A...601A..14M}. The observed variability, illustrated
in Fig.~\ref{fig:rv_hjd_200311}, indicates that the orbital period is
much longer than the time base spanned by the observations ($\sim$ 12.2\,yr). 

\subsection{Other stars}
\label{sec:SB:others}
Our single new measurement of the RV of HD~2453 is consistent with the
long-term variability reported by \citetalias{2017A&A...601A..14M},
which is 
indicative of an orbital period $\Porb\gg2850$\,d. Similarly, adding a
single new RV value for HD~12288 does not allow us to refine
the determination of its orbital elements
\citepalias{2017A&A...601A..14M}. The value that we derive for the
RV of HD~59435 appears consistent with the orbital solution of 
\cite{1999A&A...347..164W}. For HD~119027, as stated in
Sect.~\ref{sec:hd_119027}, 
comparison of the RV values of 2000--2002 with those of 1993--1996
hints at a small but 
significant decrease. Longer-term follow-up in progress should allow
the variability to be confirmed and further characterised. While the
two RV values that we measured for HD~157751 do not show a definite
variation, they are significantly
different from the value of the Gaia Data Release 3
\citep{2023A&A...674A...1G}, $\vr=-7.60\pm1.11$\,\kms. This suggests
that HD~157751 may be a spectroscopic binary.

To the limited
precision achievable in the 
determination of its RV, HD~12098 does not show any
indication of binarity. The two new RV values obtained for HD~51684
confirm the lack of variability observed in the data of
\citetalias{2017A&A...601A..14M}. Radial velocity variation of the 
order of 1\,\kms\ may occur in HD~70331 but probably reflects rotational
modulation due to surface abundance inhomogeneities. The standard
deviation of our 19 RV measurements in HD~75445, 0.13\,\kms, is
remarkably low; there is no hint of binarity. There is no evidence of
RV variations in HD~93507 either. Our two measurements of the RV of
HD~94427 do not significantly differ from the eight values reported by
\citet{2002A&A...394..151C}. The latter do not indicate any
variability. No evidence of RV variability is found for HD~126515,
HD~134214, HD~137949, or HD~177765. Our measurements of the RV of
HD~150562 do not show any significant variability either, consistent
with the fact that this star has 
been selected as an RV standard for {\em Gaia\/} Data
Release 2 \citep{2018A&A...616A...7S}. The values that we measure are
in good agreement with the value given in this reference,
$\vr=-39.519\pm0.0019$\,\kms. Our single new measurement of the RV of
HD~192678 does not show any change with respect to the values of
\citetalias{2017A&A...601A..14M}. Neither does the RV of HD~213637
show any significant variability. This is consistent with the
inclusion of this star in the catalogue of RV standards for Gaia DR2
\citep{2018A&A...616A...7S}. Within errors, the values that we
measure are consistent with the value given in the catalogue,
$\vr=5.261\pm0.099$\,\kms. We also confirm
the conclusion drawn by \citetalias{2017A&A...601A..14M}, that the RV
of HD~216018 remained constant over the time interval spanned by the
analysed observations.

The RV of HD~318107 varies with the rotation period,
$\Prot=9\fd7091$. This reflects the inhomogeneous distribution of iron
across the stellar surface, as illustrated in
Fig.~\ref{fig:rv_ew_318107}, where we compare the variation with
rotation phase of the RV derived from the analysis of the \Feline\
line and the equivalent width $W_\lambda$ of the Fe~{\sc
  ii}\,$\lambda$\,6147.7\,\AA\ line. (The latter proves more suitable
than the former for $W_\lambda$ measurements because it is not
significantly affected by blends.) The two quantities vary roughly in
phase quadrature, as expected for an approaching and receding region
of greater Fe abundance. The non-uniform distribution of the elemental
abundances over the stellar surface had already been noted by
\citet{2011A&A...535A..25B}. The Fe abundance is lowest close to the
primary maximum of the mean magnetic field modulus (see
Fig.~\ref{fig:bm_phi_318107}).

\onecolumn 

\section{Tables}
\label{sec:tables}
\begin{table*}[h!]
  \scriptsize
\caption{Magnetic field, rotation, and binarity.}
\label{tab:stars}
\begin{tabular*}{\textwidth}[]{@{}@{\extracolsep{\fill}}rrrcclcclcrlcc
@{\extracolsep{0pt}}@{}}
\hline\hline \\[-4pt]
  \multicolumn{1}{c}{HD}&\multicolumn{1}{c}{$N_{\rm
                          m}$}&\multicolumn{1}{c}{$\Bm_{\rm av}$ (G)}&$q$&\multicolumn{1}{c}{Ref.}&\multicolumn{1}{c}{$\Prot$}&\multicolumn{1}{c}{Ref.}&${\rm HJD}_0$&Phase origin&Ref.&\multicolumn{1}{c}{$N_{\rm RV}$}&\multicolumn{1}{c}{$\Porb$}&$e$&\multicolumn{1}{c}{Ref.}\\[4pt]
  \hline \\[-4pt]
  2453& 1&3666&1.0&2&518.2\,d&3&2\,448\,440.911&$\Bz$ min.&2& 1&$\gg2850$\,d&&4\\
  12098& 4&{\it 7072}&{\it 1.5}&1&$5\fd460$&5&&&& 4\\
  12288& 1&8065&1.1&2&$34\fd993$&6&2\,451\,131.900&$\Bz$ max.&2& 1&$1546\fd1$&0.342&4\\
  29578&25&3820&2.2&1&3950\,d&1&2\,449\,600.000&$\Bm$ min.&1&31&$926\fd5$&0.387&1\\
  47103&18&{\it 17168}&{\it 1.1}&1&$\gg6.8$\,yr&1&&&& 7&SB&&4\\
  51684&17&6053&1.1&1&366\,d&2&2\,449\,960.000&$\Bm$ max.&1&12\\
  55719&92&{\it 6027}&{\it 1.1}&1&$\gg30$\,yr&1&&&&111&46\fd31914&0.1451&1\\
  59435& 1&3059&1.9&4&1360\,d&7&2\,450\,580.000&$\Bm$ max.&7& 1&$1386\fd1$&0.285&7\\
  61468&16&6768&1.4&1&323\,d&1&2\,450\,058.500&$\Bm$ max.&2&21&$27\fd2729$&0.1730&1\\
  70331& 1&{\it 12392}&{\it 1.1}&4&$1\fd9909$?&4&&&&39\\
  75445&21&{\it 2847}&{\it 1.1}&1&$\gg13$\,yr?&1&&&&19\\
  81009&84&8422&1.34&1&$33\fd987$&3&2\,448\,646.878&$v$ min.&3&89&11998\,d&0.736&1\\
  93507&63&7060&1.08&1&562\,d&2&2\,448\,965.000&$\Bm$ max.&1&47\\
  94427& 2&{\it 1686}&{\it 1.62}&1&$1\fd9625$&8&&&&2\\
  94660&50&6239&1.05&1&2830\,d&2&2\,447\,000.000&$\Bm$ min.&4&71&$848\fd94$&0.4443&1\\
  110066&23&{\it 4037}&{\it 1.08}&1&$\gg33$\,yr?&&&&&17&SB?\\
  116114&68&{\it 6064}&{\it 1.08}&1&$\ga48$\,yr&2&&&&70&3956\,d&0.648&1\\
  119027&15&{\it 3033}&{\it 1.16}&1&$\gg9$\,yr&1&&&&12&SB&&1\\
  126515&65&12499&1.73&1&$129\fd95$&9&2\,437\,015.000&$\Bm$ max.&4&41\\
  134214&50&{\it 2935}&{\it 1.08}&1&$\gg23.5$\,yr?&&&&&50\\
  137949&49&{\it 4779}&{\it 1.03}&1&$\gg19$\,yr?&&&&&46\\
  142070&44&4653&1.06&1&$3\fd37204$&1&2\,449\,878.200&$\Bm$ max.&2&74&4598\,d&0.651&1\\
  144897&60&8995&1.14&1&$48\fd60$&2&2\,447\,000.000&$\Bm$ min.&1&43&3310\,d&0.667&1\\
  150562&29&{\it 4926}&{\it 1.05}&1&$>21$\,yr&1&&&&15\\
  157751& 3&{\it 5782}&{\it 1.30}&1&$2\fd70134$&10&&&&2&SB?&&1\\
  318107&53&14258&1.15&1&$9\fd7091$&1&2\,448\,800.000&primary $\Bm$ max.&2&38\\
  165474&57&6605&1.07&1&9750\,d&1&2\,447\,515.000&$\Bm$ min.&1&63&$>12000$\,d&&1\\
  177765&11&{\it 3392}&{\it 1.09}&1&$\ga50$\,yr&&&&1&11\\
  187474&54&5395&1.27&1&2345\,d&11&2\,446\,766.000&$\Bz$ pos. extr.&11&115&$689\fd67$&0.482&1\\
  192678& 1&4624&1.05&2&$6\fd4199$&2&2\,449\,112.760&$\Bm$ max.&2& 1\\
  335238&22&9471&1.52&1&$48\fd990$&1&2\,447\,450.000&secondary $\Bm$ max.&1&22&long?&&1\\
  200311&31&8279&1.18&1&$52\fd0084$&12&2\,445\,407.513&$\Bz$ max.&12&31&$\gg12.5$\,yr&&1\\
  213637&10&{\it 5323}&{\it 1.11}&1&$>8$\,yr&1&&&8\\ 
  216018&42&{\it 5921}&{\it 1.07}&1&$>26$\,yr&1&&&&26\\[4pt]
  \hline
\end{tabular*}
\tablefoot{
The stars are identified in Col.~1 by their HD numbers; they are
listed in order of increasing right ascension. Column~2 gives the
number $N_{\rm m}$ of mean magnetic field modulus determinations used in the
present analysis. This includes values from the literature when they
were considered for characterisation of the $\Bm$ variation curve. For
those stars for which we added a single new measurement that proved
consistent with the conclusions of \citetalias{2022MNRAS.514.3485G} or
\citetalias{2017A&A...601A..14M},
$N_{\rm m}=1$ appears in this column. For the known period stars for
which a fitted $\Bm$ phase curve has been computed, the average
value $\Bm_{\rm av}$ is the $M_0$ parameter 
of fit of the mean magnetic field modulus by a cosine wave or the
superposition of a cosine and its first harmonic (see
Table~\ref{tab:mfit}); the ratio $q$ of the $\Bm$ extrema is computed
by dividing the maximum of the fitted curve by its minimum. 
When the rotation period is unknown, the
average of the analysed $\Bm$ values and the ratio of the largest one
to the smallest one are used instead. They appear in
italics to stress the fact that they may not accurately represent the
strength and variation amplitude of the mean magnetic field modulus
over a rotation cycle. The reference from which the values of
$\Bm_{\rm av}$ and $q$ are derived is identified in Col.~5. Columns~6
and 7 give the adopted value of the rotation period and its
source. Period lower limits followed by a question mark distinguish
those stars for which no definite variations have been observed yet,
which leaves open the alternative possibility that they may be seen
rotation pole-on and have shorter periods. Accordingly, no
  reference was given for the period. For the stars for which a
least-squares fit of the 
$\Bm$ variation curve was computed, the date of the phase origin, the
property to which it refers, and the corresponding reference are
listed in Cols.~8 to 10. The number $N_{\rm RV}$ of RV measurements
that we analysed is given in Col.~11; it includes data from the
literature when we used them for computation of spectroscopic binary
orbital 
elements. For the eleven systems for which these elements could be
determined (nine derived or refined in the present study, the other
two extracted from the literature -- see references in Col.~14),
the rotation period appears in Col.~12 and the eccentricity in Col.~13.}  
\tablebib{  1:~This work; 2:~\citetalias{2022MNRAS.514.3485G}; 3:~\cite{2017PASP..129j4203P}; 4:~\citetalias{2017A&A...601A..14M}; 5:~\citet{2005A&A...429L..55R}; 6:~\citet{2000A&A...355.1080W}; 7:~\citet{1999A&A...347..164W}; 8:~\citet{2007A&A...475.1053A}; 9:~\citetalias{1997A&AS..123..353M}; 10:~\citet{2023ApJ...955..123S}; 11:~\citet{1991A&AS...89..121M}; 12:~\citet{1997MNRAS.292..748W} }
\end{table*}

\begin{table*}[h!]
\caption{Variation of the mean magnetic field modulus: least-squares
  fit parameters.}
\label{tab:mfit}
\begin{tabular*}{\textwidth}[]{@{}@{\extracolsep{\fill}}rcccccrcc
@{\extracolsep{0pt}}@{}}
\hline\hline \\[-4pt]
\multicolumn{1}{c}{HD/HDE}&$M_0\pm\sigma$&$M_1\pm\sigma$
&$\phi_{M_1}\pm\sigma$&$M_2\pm\sigma$
&$\phi_{M_2}\pm\sigma$&\multicolumn{1}{c}{$\nu$}&$\chi^2/\nu$&$R$\\
&(G)&(G)&&(G)\\[4pt]
\hline \\[-4pt]
 29578&$\fo3820\pm15$&$1476\pm17$&$0.569\pm0.002$&$290\pm14$&$0.259\pm0.013$&20&1.5&1.00\\
 51684&$\fo6053\pm\fo5$&$\fo265\pm\fo8$&$0.992\pm0.005$&&&14&0.8&0.99\\
 61468&$\fo6768\pm27$&$1162\pm36$&$0.998\pm0.005$&&&13&7.6&0.99\\
 81009&$\fo8422\pm\fo8$&$1238\pm10$&$0.964\pm0.001$&$\fo61\pm\fo9$&$0.515\pm0.033$&79&0.7&1.00\\
 93507&$\fo7060\pm\fo8$&$\fo262\pm13$&$0.013\pm0.007$&$\fo62\pm13$&$0.372\pm0.032$&58&1.2&0.96\\
 94660&$\fo6239\pm\fo5$&$\fo164\pm\fo7$&$0.459\pm0.007$&&&47&1.8&0.96\\
126515&$12499\pm15$&$3357\pm24$&$0.017\pm0.001$&$220\pm23$&$0.893\pm0.016$&60&2.6&1.00\\
142070&$\fo4653\pm13$&$\fo142\pm20$&$0.897\pm0.022$&$217\pm20$&$0.004\pm0.014$&39&2.3&0.94\\
144897&$\fo8995\pm\fo8$&$\fo582\pm11$&$0.468\pm0.003$&$107\pm11$&$0.845\pm0.017$&55&1.4&0.99\\
318107&$14258\pm61$&$\fo980\pm88$&$0.970\pm0.014$&$634\pm85$&$0.929\pm0.023$&48&2.0&0.90\\
165474&$\fo6605\pm\fo9$&$\fo210\pm17$&$0.503\pm0.008$&&&54&2.7&0.88\\  
187474&$\fo5395\pm\fo7$&$\fo644\pm10$&$0.453\pm0.003$&$224\pm11$&$0.954\pm0.007$&49&4.0&0.99\\
335238&$\fo9471\pm45$&$1953\pm67$&$0.491\pm0.005$&$906\pm58$&$0.980\pm0.011$&36&3.4&0.99\\
200311&$\fo8279\pm36$&$\fo697\pm45$&$0.051\pm0.013$&$484\pm45$&$0.084\pm0.015$&26&0.3&0.97\\[4pt]
\hline
\end{tabular*}
\tablefoot{The HD numbers of each star are listed in Col.~1. Columns~2
  to 6 present the fit coefficients, as defined in Eq.~(\ref{eq:fit1})
  or Eq.~(\ref{eq:fit2}).  Columns~7 to 9 give the number of degrees
  of freedom about the fit $\nu$, the reduced $\chi^2$ of the fit
  $\chi^2/\nu$, and the multiple correlation coefficient $R$.}
\end{table*}

\begin{table*}[h!]
\caption{New or revised orbital elements of spectroscopic
  binaries. 
}
\label{tab:orbits}
\begin{tabular*}{\textwidth}[]{@{}@{\extracolsep{\fill}}rrrrrrrrrr
@{\extracolsep{0pt}}@{}}
\hline\hline \\[-4pt]
\multicolumn{1}{c}{HD}&\multicolumn{1}{c}{$P_{\rm orb}$}&\multicolumn{1}{c}{$T_0$}&\multicolumn{1}{c}{$e$}&\multicolumn{1}{c}{$V_0$}&\multicolumn{1}{c}{$\omega$}&\multicolumn{1}{c}{$K$}&\multicolumn{1}{c}{$f(M)$}&\multicolumn{1}{c}{$a\,\sin
  i$}&\multicolumn{1}{c}{$N$}\\
&\multicolumn{1}{c}{(d)}&\multicolumn{1}{c}{(HJD)}&&\multicolumn{1}{c}{(km\,s$^{-1}$)}&\multicolumn{1}{c}{($^\circ$)}&\multicolumn{1}{c}{(km\,s$^{-1}$)}&$(M_{\sun})$&\multicolumn{1}{c}{($10^6$\,km)}&\\[4pt]
  \hline \\[-4pt]
29578&926.5&2449151.6&0.387&15.70&226.9&9.71&0.0690&114.1&31\\
                      &1.6&5.2&0.014&0.11&2.8&0.23&0.0051&2.8&\\[4pt]   
55719&46.31914&2441674.05&0.1451&-8.13&205.1&47.37&0.494&29.85&111\\
                      &0.00065&0.43&0.0080&0.28&3.4&0.40&0.013&0.25\\[4pt]  
61468&27.2729&2445594.21&0.1730&31.82&182.2&36.48&0.1310&13.48&21\\
                      &0.0015&0.30&0.0058&0.18&2.4&0.29&0.0031&0.11\\[4pt]
81009&11998&2450784&0.736&26.30&147.4&7.02&0.1332&784&89\\
                      &618&22&0.016&0.10&1.6&0.16&0.0138&48\\[4pt]
94660&848.94&2445627.8&0.4443&18.49&264.0&17.88&0.3611&187.0&71\\
                      &0.12&1.6&0.0040&0.06&0.7&0.09&0.0064&1.0&\\[4pt]
116114&3956&2448179&0.648&5.03&170.5&5.83&0.0359&242&70\\
                      &15&19&0.024&0.09&1.5&0.41&0.0081&18&\\[4pt]
142070&4598&2449753&0.651&$-$8.63&218.7&1.73&0.00107&82.8&74\\
                      &24&17&0.023&0.06&4.6&0.07&0.00016&4.1&\\[4pt]
144897&3310&2451501&0.667&2.92&353.6&9.54&0.1230&323&43\\
                      &14&13&0.029&0.11&2.2&0.42&0.0207&18&\\[4pt]
187474&689.67&2435139.0&0.482&$-$1.76&7.0&11.05&0.0648&91.8&115\\
  &0.10&2.3&0.009&0.07&1.4&0.14&0.0028&1.3\\[4pt]
\hline
\end{tabular*}
\tablefoot{The columns give, in order, the HD number of the star, the
  orbital period, the time of the periastron passage, the orbital
  eccentricity, the systemic RV, the longitude of the
  orbital periastron, the velocity semi-amplitude, the mass function,
  the projected semi-major axis of the orbit, and the number of RV
  measurements used for computation of the orbital elements. The
  second row for each entry gives its standard deviation.}
\end{table*}

\end{appendix}

\end{document}